\documentclass[11pt, showpacs, amsmath, amssymb, aps, prd]{revtex4-2}
\usepackage{graphicx}
\usepackage{dcolumn}
\usepackage{bm}
\usepackage{hyperref}
\usepackage{color,xcolor}
\usepackage{bigints}
\usepackage{mathtools} 

\allowdisplaybreaks

\begin{document}

\title{Multi-throat generalized Ellis--Bronnikov wormholes with asymptotic string-cloud behavior}

\author{M. S. Cunha}\email{marcony.cunha@uece.br}
\affiliation{Universidade Estadual do Ceará (UECE), Centro de Ciências e Tecnologia (CCT), CEP 60714-903, Fortaleza-CE, Brazil}
\author{C. R. Muniz}\email{celio.muniz@uece.br}\affiliation{Universidade Estadual do Ceará (UECE), Faculdade de Educação, Ciências e Letras de Iguatu (FECLI), CEP 63502-253, Iguatu-CE, Brazil.}
\author{V. H. U. Borralho}\email{victorborralho@fisica.ufc.br} \author{G. Alencar}\email{geova@fisica.ufc.br} 
\affiliation{Departamento de Física, Universidade Federal do Ceará (UFC), Caixa Postal 6030, Campus do Pici, 60455-760 Fortaleza, Ceará, Brazil}

\begin{abstract}
We study an ultrastatic, spherically symmetric family of traversable wormholes obtained by locally deforming a generalized Ellis--Bronnikov areal-radius profile in a spacetime with an asymptotic angular deficit. Depending on the deformation parameters, the same analytic family contains one, two, three, or four throats. The corresponding bifurcation curves are obtained in dimensionless form, and the resulting geometries are regular and asymptotically conical. We also reconstruct a source in General Relativity using a canonical scalar, a shift-symmetric phantom scalar, and a magnetic nonlinear electromagnetic field coupled to the canonical scalar. In the reference multi-throat geometry, the coupling removes a branch obstruction of an uncoupled single-valued nonlinear electrodynamics (NED); the same obstruction occurs whenever equal values of the magnetic invariant require different values of $\mathcal G_F$. For $\mathcal G(F)=F+\lambda\sqrt F$, the NED sector carries the leading asymptotic stress, which scales as $r^{-2}$ and approaches the string-cloud equation of state. The total stress tensor satisfies pointwise energy conditions in finite belly-centered regions of representative two- and four-throat solutions, while the throats violate the radial null energy condition (NEC) and the violation extends into integrable asymptotic tails. The Nandi--Zhang--Kumar covariant volume integral of $\rho+p_r$ is quadratic in the deformation amplitude; its zero is distinct from the bifurcation curves. Away from saddle-node boundaries, each throat supports a radially unstable background-metric light ring, and each intermediate belly supports a radially stable one, with an internally trapped null family in the double-throat case. These results place the topology, source reconstruction, integrated NEC content, and background-metric circular-orbit structure in a common parameter-space description.
\end{abstract}

\maketitle

\section{Introduction}

Traversable wormholes provide some of the most direct laboratories for testing the interplay between spacetime topology, energy conditions, and matter models in General Relativity. The modern formulation by Morris and Thorne established the geometric requirements for a static, spherically symmetric, horizonless tunnel and made explicit the relation between the flare-out condition and violations of the null energy condition (NEC) near a regular throat \cite{MorrisThorne1988}. The broader geometric and causal structure of Lorentzian wormholes was subsequently systematized by Visser \cite{Visser1995}, while Hochberg and Visser formulated the throat conditions in a covariant language and clarified the central role played by null convergence and energy-condition violation \cite{HochbergVisser1998}. These results explain why the source problem remains one of the central obstacles in classical wormhole physics: within Einstein gravity, a regular traversable throat requires a stress tensor that is not everywhere compatible with the usual pointwise energy conditions.

This observation motivated two complementary lines of research. One seeks to modify the gravitational sector, transferring the effective NEC violation to higher-curvature or nonminimal terms; examples include higher-derivative gravity, scalar--Gauss--Bonnet models, and a broad range of extended theories \cite{Antoniou2020,Radhakrishnan2024}. The other retains standard General Relativity and asks whether the exotic component can be localized, diluted, or separated from regions supported by ordinary matter. Thin-shell constructions are an important realization of the latter strategy \cite{Lobo2025}. Closely related is the quantitative question of how much energy-condition-violating matter a wormhole contains. Visser, Kar, and Dadhich introduced an areal-coordinate volume integral and showed that suitable geometries can be supported by arbitrarily small integrated violations \cite{VisserEC}. Nandi, Zhang, and Kumar later proposed an exact covariant volume-integral quantifier based on the measure $\sqrt{-g}\,d^3x$ \cite{NandiEC}. In the ultrastatic proper-radial gauge used below, the latter measure reduces directly to $4\pi r^2d\ell$. These integrals are algebraic: NEC-satisfying and NEC-violating regions can cancel, so their sign should not automatically be interpreted as a positive measure of the total exotic content.

The Ellis--Bronnikov geometry is the canonical exact example of a traversable wormhole in Einstein gravity. It was obtained independently in scalar-field settings by Ellis and Bronnikov and is supported by a massless scalar with reversed kinetic sign \cite{Ellis1973,Bronnikov1973}. Its ultrastatic form, regularity, geodesic completeness, and analytic simplicity have made it a standard benchmark for both classical and observational studies. Related one-parameter ultrastatic wormhole families were explored by Kar, Minwalla, Mishra, and Sahdev in the context of wave propagation \cite{Kar1995}; the form now commonly referred to as the generalized Ellis--Bronnikov (GEB) geometry introduces an even power controlling the local flattening or steepness of the areal-radius profile. GEB wormholes have since been examined in modified gravity \cite{Sokoliuk2022} and, more recently, their matter sources have been reconstructed explicitly in General Relativity using scalar and electromagnetic sectors \cite{CrispimGEB2026}. The generalized profile is especially useful because it separates a robust ultrastatic background from the detailed geometry of the central region.

An important development beyond the standard Ellis picture is the appearance of wormholes with more than one minimal-area surface. Double-throat configurations, typically separated by an equator or ``belly,'' arise in several independent systems. They were found in wormholes threaded by chiral fields, where increasing the matter coupling can replace a single throat by two throats with an intermediate belly \cite{Kunz2013Chiral}; analogous structures occur in non-Abelian Einstein--Yang--Mills--Higgs wormholes \cite{Chew2020}, in Einstein--scalar--Gauss--Bonnet models with scalar self-interactions \cite{Ibadov2020}, and in Einstein--vector--Gauss--Bonnet theory \cite{Barton2022}. More general multiple-throat/anti-throat geometries and their field sources were analyzed in Ref.~\cite{CrispimMulti2025}. Recently, a systematic two-throat construction in ordinary General Relativity was presented by Cataldo, Cid, and Labra\~na \cite{Cataldo2026}, and double-throat phases have also appeared in self-gravitating axion configurations with wormhole topology \cite{Hao2026}. Taken together, these results show that multiple extrema of the areal radius are not tied to a single theory or matter model; they are a recurring geometric response of gravitating systems with sufficiently structured sources.

A preliminary version of this program was reported by some of us in Ref.~\cite{Amaral2026Preliminary}. That work used a localized deformation of the standard Ellis--Bronnikov profile and focused on the double-throat sector with string-cloud asymptotics. Here the undeformed geometry is replaced by the generalized Ellis--Bronnikov family and the extremum problem is solved globally, revealing one-, two-, three-, and four-throat phases. The present analysis also adds the two-scalar scalar-coupled NED reconstruction, the branch obstruction of an uncoupled single-valued NED, the exact sign transition of the global radial-NEC volume integral, and the background-metric light-ring classification. Ref.~\cite{Amaral2026Preliminary} is therefore the restricted precursor of the broader construction developed here.

The field-theoretic realization of such geometries raises an additional difficulty. A single scalar with a kinetic prefactor that changes sign can interpolate between canonical and phantom regions, an idea developed in the ``trapped ghost'' construction \cite{BronnikovSushkov2010}. However, the sign-changing kinetic coefficient becomes degenerate at the transition surface and can generate singular structures in the radial perturbation potential, as reviewed in Ref.~\cite{BronnikovReview2018}. An alternative is to use distinct scalar degrees of freedom with fixed kinetic signs. Bronnikov and Korolyov employed canonical and phantom scalars together with an electromagnetic sector to construct regular wormholes and black-universe solutions while making the phantom contribution rapidly decay in the weak-field region \cite{BronnikovKorolyov2015}. Scalar fields combined with nonlinear electrodynamics (NED) have also proved useful for reconstructing generalized and multi-throat wormhole geometries \cite{CrispimMulti2025,CrispimGEB2026,SuHao2025}. These precedents motivate the two-scalar plus NED realization adopted in the present work.

The asymptotic sector offers a further physical guide. Letelier's string-cloud model describes an ensemble of radially oriented strings and yields a characteristic anisotropic stress tensor with density falling as the inverse square of the areal radius \cite{Letelier1979}. Spherically symmetric string-cloud geometries carry an angular-deficit structure closely analogous, at the level of the far-field metric, to the deficit produced by a global monopole \cite{BarriolaVilenkin1989}; more recent generalizations of string clouds have explored related matter constructions \cite{Alencar2025Cloud}. An asymptotically conical wormhole supported by a source that tends to this radial string equation of state therefore has a natural interpretation that is more specific than a generic anisotropic fluid. The geometric deficit already fixes the total leading $r^{-2}$ stress required asymptotically. What is model dependent is how that stress is distributed among the reconstructed fields: with the specific choice $\mathcal G(F)=F+\lambda\sqrt F$ adopted below, the leading $r^{-2}$ contribution is carried entirely by the magnetic NED sector, while the scalar contributions are subleading.

Multi-throat structure is also expected to leave a direct imprint on null and timelike motion. The Ellis wormhole already possesses an unstable circular photon orbit at its throat, and its imaging properties have been studied in optically thin environments \cite{OhgamiSakai2015}. More complicated wormholes can support multiple photon rings and qualitatively different shadow morphologies, as illustrated by reflection-asymmetric double-shadow configurations \cite{Wielgus2020} and by studies of light rings on different sides of asymmetric Ellis--Bronnikov wormholes \cite{HuangKunz2023}. In multi-throat geometries the issue is particularly transparent in proper-radial coordinates: every extremum of the areal radius is a candidate circular orbit, but its stability depends on whether the extremum is a throat or a belly. This makes the optical sector a useful diagnostic of the same topological bifurcations that govern the embedding geometry.

Here we place these ingredients in a single GEB-based family. A localized even deformation and an overall deficit parameter are added to the areal-radius profile, and the full extremum problem is solved in the proper-radial coordinate. The phase diagram contains one-, two-, three-, and four-throat sectors. Noncentral throat--belly pairs appear or disappear through saddle-node folds, while the symmetric central transition may instead be controlled by an even-order degeneracy. The global analysis is essential because the local character of the center does not, by itself, determine the total number of throats.

The source is reconstructed directly in Einstein gravity with two scalar fields of fixed kinetic sign and a magnetic NED sector. The phantom scalar is shift symmetric, while the canonical scalar carries the potential and the scalar-dependent electromagnetic coupling. In the reference multi-throat geometry, the nonmonotonic areal radius creates a branch obstruction for an uncoupled NED: different radial positions can have the same magnetic invariant but require different values of $\mathcal G_F$. Whenever $Q(r)$ is multivalued, the monotonic scalar supplies an additional label that removes this ambiguity. In the sign-controlled reference sector the relevant electromagnetic derivatives have the Maxwell sign. The scalar terms become subleading at large radius, and for the chosen NED model the leading stress approaches the string-cloud form associated with the conical asymptotics.

The energy conditions follow directly from the local concavity of the areal radius. In representative two- and four-throat solutions, finite regions centered on bellies satisfy the WEC, DEC, and SEC for the total stress tensor, whereas the throat sectors violate the radial NEC. The identity $\rho+p_r+2p_t=0$ leaves no additional SEC inequality in the ultrastatic class. Although the radial-NEC violation reaches the asymptotic regions, its tails are integrable, and the Nandi--Zhang--Kumar volume integral changes sign along a curve that is independent of the local throat condition. The same throat--belly sequence controls the background-metric circular orbits: away from bifurcation curves, throats carry unstable light rings and intermediate bellies stable ones. The double-throat reference solution also contains an internal trapped null family that cannot be fed by conservative rays arriving from infinity.

The paper is organized as follows. Section~II introduces the geometry, derives the global bifurcation equations, classifies the one- through four-throat sectors, and analyzes embeddings, regularity, and asymptotics. Section~III reconstructs the two-scalar plus NED source and identifies the asymptotic string-cloud regime. Section~IV studies the pointwise energy conditions and the covariant Nandi--Zhang--Kumar integral of the radial-NEC content. Section~V develops the null and timelike geodesic structure and the associated light-ring counting. We close with a summary of the main results, limitations, and directions for perturbative and observational studies.

\section{Geometry and Topological Bifurcations}\label{field_eqs}

This section introduces the ultrastatic wormhole geometry and establishes the global classification of the extrema of the areal radius. In contrast with a purely local analysis based only on $r''(0)$, the number of throats is determined by the complete set of roots of $r'(\ell)=0$. This distinction is essential for the family considered below, which admits one-, two-, three-, and four-throat configurations in different regions of parameter space.

\subsection{Metric and Einstein equations}

We consider a static and spherically symmetric spacetime written in proper radial distance coordinates,
\begin{equation}
    ds^2=-e^{2\Phi(\ell)}dt^2+d\ell^2+r^2(\ell)d\Omega^2,
    \label{eq:metric}
\end{equation}
where $\ell\in(-\infty,+\infty)$ and $r(\ell)>0$ is the areal radius. A throat is a local minimum of $r(\ell)$, whereas a local maximum separating neighboring throats will be referred to as a \textit{belly}. Throughout this work we restrict the analysis to the ultrastatic, zero-redshift class,
\begin{equation}
    \Phi(\ell)=0,
\end{equation}
so that $g_{tt}=-1$.

We write Einstein's equations as
\begin{equation}
    G_{\mu\nu}=\kappa T_{\mu\nu},
\end{equation}
with $\kappa=8\pi G$ ($\kappa=8\pi$ in units $G=1$), and use the anisotropic form
\begin{equation}
    T^{\mu}{}_{\nu}=\mathrm{diag}(-\rho,p_r,p_t,p_t).
\end{equation}
The independent field equations are
\begin{align}
    \kappa\rho(\ell)&=\frac{1-r'(\ell)^2-2r(\ell)r''(\ell)}{r^2(\ell)},
    \label{eq:rho_general}\\
    \kappa p_r(\ell)&=\frac{r'(\ell)^2-1}{r^2(\ell)},
    \label{eq:pr_general}\\
    \kappa p_t(\ell)&=\frac{r''(\ell)}{r(\ell)}.
    \label{eq:pt_general}
\end{align}
In particular,
\begin{equation}
    \kappa(\rho+p_r)=-\frac{2r''}{r},
    \label{eq:radial_nec_geometry}
\end{equation}
which will be used later to relate the radial null energy condition directly to the local concavity of the areal-radius profile.

\subsection{Localized deformation of the generalized Ellis geometry}

We take
\begin{equation}
    r(\ell)=\alpha\left(\ell^m+r_0^m\right)^{1/m}
    +\frac{A}{1+(\ell/b)^n},
    \label{eq:master_function}
\end{equation}
where $m,n\geq2$ are even integers, $r_0>0$, $b>0$, $A\geq0$, and $0<\alpha<1$. The first term is the generalized Ellis--Bronnikov profile multiplied by the asymptotic conical factor $\alpha$, while the second term is an even, localized rational deformation. The parameter $r_0$ sets the length scale of the undeformed geometry; $b$ controls the width of the deformation; and $A$ controls its amplitude. The asymptotic interpretation of $\alpha$ and the matter source generating the conical sector are discussed below.

It is convenient to introduce dimensionless quantities
\begin{equation}
    x=\frac{\ell}{r_0},\qquad
    \beta=\frac{b}{r_0},\qquad
    a=\frac{A}{\alpha r_0},\qquad
    \mathcal R(x)=\frac{r(\ell)}{\alpha r_0}.
    \label{eq:dimensionless_variables}
\end{equation}
Equation~\eqref{eq:master_function} then becomes
\begin{equation}
    \mathcal R(x)=\left(1+x^m\right)^{1/m}
    +\frac{a}{1+(x/\beta)^n}.
    \label{eq:dimensionless_profile}
\end{equation}
Apart from the overall scales, the topology is controlled by $(m,n)$ and the two dimensionless parameters $(a,\beta)$. We focus on $A\geq0$, hence $a\geq0$. A formal extension to $a<0$ is topologically trivial: for $x>0$ the undeformed derivative is positive and the deformation term contributes with the same sign, so $\mathcal R'(x)>0$ everywhere. The entire negative-amplitude branch is therefore $1T$.

The derivative of the profile is
\begin{equation}
\mathcal R'(x)=
\frac{x^{m-1}}{(1+x^m)^{(m-1)/m}}
-\frac{a n\beta^n x^{n-1}}{(\beta^n+x^n)^2}.
\label{eq:profile_derivative}
\end{equation}
Because the profile is even, $x=0$ is always an extremum. The noncentral extrema, $x>0$, satisfy
\begin{equation}
    a=\mathcal A(x)\equiv
    \frac{x^{m-n}(\beta^n+x^n)^2}
    {n\beta^n(1+x^m)^{(m-1)/m}}.
    \label{eq:a_of_x}
\end{equation}
This representation is useful because, for fixed $(m,n,\beta)$, the intersections of a horizontal line $a=\mathrm{const.}$ with $\mathcal A(x)$ count all noncentral extrema of $r(\ell)$ on one side of the symmetric geometry.

\subsection{Local character of the central extremum}

Near $\ell=0$, Eq.~\eqref{eq:master_function} gives
\begin{equation}
\begin{split}
    r(\ell)=&\,\alpha r_0+A
    +\frac{\alpha}{m r_0^{m-1}}\ell^m
    -\frac{A}{b^n}\ell^n\\
    &+\mathcal O(\ell^{2m},\ell^{2n}).
\end{split}
\label{eq:central_expansion}
\end{equation}
Consequently, the sign of $r''(0)$ is sufficient only when the first nonvanishing correction is quadratic. For $m,n>2$, one must instead inspect the first nonzero even derivative.

Three qualitatively distinct cases follow immediately:

\begin{itemize}
    \item \textbf{$n>m$.} The positive $\ell^m$ term dominates, and $\ell=0$ is a (possibly degenerate) local minimum. Hence the center is a throat.

    \item \textbf{$m>n$.} The negative $\ell^n$ term dominates, and $\ell=0$ is a (possibly degenerate) local maximum. Hence the center is a belly for every $A>0$.

    \item \textbf{$m=n$.} The leading correction is
    \begin{equation}
        r(\ell)-r(0)=
        \left(\frac{\alpha}{m r_0^{m-1}}-\frac{A}{b^m}\right)\ell^m+
        \mathcal O(\ell^{2m}),
    \end{equation}
    so that the central extremum changes character at
    \begin{equation}
        A_c=\frac{\alpha b^m}{m r_0^{m-1}},
        \qquad
        a_c=\frac{\beta^m}{m}.
        \label{eq:critical_A_equal_mn}
    \end{equation}
    The center is a throat for $A<A_c$ and a belly for $A>A_c$; $A=A_c$ is a degenerate critical configuration.
\end{itemize}

For $m=n=2$, Eq.~\eqref{eq:critical_A_equal_mn} reduces to the familiar quadratic condition $A_c=\alpha b^2/(2r_0)$. For $m=n>2$, however, the transition is governed by the first nonvanishing derivative of order $m$ rather than by $r''(0)$.

\subsection{Global bifurcation equation and number of throats}

The local character of the center does not by itself determine the global number of throats. Additional extrema are created or annihilated when two roots of $r'(\ell)=0$ merge, i.e., when
\begin{equation}
    r'(\ell_*)=0,
    \qquad
    r''(\ell_*)=0.
    \label{eq:saddle_node_conditions}
\end{equation}
Equivalently, these bifurcations occur at stationary points of $\mathcal A(x)$. Differentiating Eq.~\eqref{eq:a_of_x} yields the compact condition
\begin{equation}
\mathcal B(x)\equiv
\beta^n\left[(m-n)-(n-1)x^m\right]
+(m+n)x^n
+(n+1)x^{m+n}=0.
\label{eq:bifurcation_equation}
\end{equation}
For every positive solution $x_*$ of Eq.~\eqref{eq:bifurcation_equation}, the corresponding critical amplitude is
\begin{equation}
    a_*=\mathcal A(x_*),
    \qquad
    A_*=\alpha r_0 a_*.
    \label{eq:bifurcation_amplitude}
\end{equation}
Equation~\eqref{eq:bifurcation_equation} also fixes the type of critical point. A generic noncentral bifurcation has $r'=r''=0$ and $r'''\neq0$, so the throat and belly meet at a stationary inflection (a saddle-node fold). At the codimension-two onset where two fold branches merge, additional derivatives may vanish and the local expansion must be continued to the first nonzero term. The symmetric central transition is different: there the leading nonzero correction can have even order and produce a degenerate extremum.

\paragraph{$n>m$.}
In this case $\mathcal A(x)\to\infty$ both as $x\to0^+$ and as $x\to\infty$. Moreover, when the powers in Eq.~\eqref{eq:bifurcation_equation} are ordered increasingly, the coefficients have exactly one sign change. Since $\mathcal B(0)<0$ and $\mathcal B(x)\to+\infty$, there is exactly one positive solution $x_*$. Therefore $\mathcal A(x)$ has a unique minimum. The topology is
\begin{equation}
    a<a_*:\ 1T,
    \qquad
    a>a_*:\ 3T,
    \label{eq:n_gt_m_classification}
\end{equation}
where $T$ denotes a throat. At $a=a_*$ the throat--belly pair coalesces at the noncentral saddle-node fold described above; exactly at the fold the coalesced stationary point is an inflection rather than a throat or belly. Thus the three-throat configuration in the sector $n>m$ is not automatic: it appears only above a finite threshold.

\paragraph{$m=n$.}
Equation~\eqref{eq:bifurcation_equation} factorizes as
\begin{equation}
    \mathcal B(x)=x^m\left[
    2m-(m-1)\beta^m+(m+1)x^m
    \right].
    \label{eq:equal_mn_bifurcation}
\end{equation}
A noncentral stationary point of $\mathcal A(x)$ exists only when
\begin{equation}
    \beta>\beta_c,
    \qquad
    \beta_c=\left(\frac{2m}{m-1}\right)^{1/m}.
    \label{eq:beta_c_equal_mn}
\end{equation}
If $\beta\leq\beta_c$, $\mathcal A(x)$ is monotonic and the central bifurcation at $a=a_c$ directly connects the one- and two-throat sectors,
\begin{equation}
    1T\ \longrightarrow\ 2T.
\end{equation}
For $\beta>\beta_c$, the noncentral minimum occurs at
\begin{equation}
    x_*^m=\frac{(m-1)\beta^m-2m}{m+1},
    \label{eq:xstar_equal_mn}
\end{equation}
with $a_*=\mathcal A(x_*)<a_c$. The sequence as $a$ increases is then
\begin{equation}
    1T\ \longrightarrow\ 3T\ \longrightarrow\ 2T,
    \label{eq:equal_mn_topology_sequence}
\end{equation}
with the three-throat window given by $a_*<a<a_c$.

\paragraph{$m>n$.}
The undeformed boundary $a=0$ must be distinguished from the genuinely deformed sector. At $a=0$ the rational term vanishes and the geometry reduces to the single-throat generalized Ellis profile. For every $a>0$, however, the leading central correction is the negative $x^n$ term, so the center is a belly. In this sector $\mathcal A(x)\to0$ as $x\to0^+$ and $\mathcal A(x)\to\infty$ as $x\to\infty$. When the powers in Eq.~\eqref{eq:bifurcation_equation} are ordered increasingly, the coefficients have the sign pattern $+,+,-,+$; hence $\mathcal B(x)$ has either no positive root or two positive roots. If it has no positive root, $\mathcal A(x)$ is monotonic and every $a>0$ produces one noncentral throat on each side of the center, i.e., a two-throat geometry. If two roots $x_-<x_+$ exist, $\mathcal A(x)$ develops a local maximum and a local minimum,
\begin{equation}
    a_{\max}=\mathcal A(x_-),
    \qquad
    a_{\min}=\mathcal A(x_+),
\end{equation}
and the interval
\begin{equation}
    a_{\min}<a<a_{\max}
    \label{eq:four_throat_window}
\end{equation}
contains three noncentral extrema for $x>0$. Their mirrored partners generate a four-throat geometry. Thus, for $a>0$, the sequence is
\begin{equation}
    2T\ \longrightarrow\ 4T\ \longrightarrow\ 2T,
    \label{eq:m_gt_n_topology_sequence}
\end{equation}
while the boundary $a=0$ remains $1T$.

The opening of the four-throat window can be obtained analytically. At the critical point the two positive roots of $\mathcal B$ merge, so $\mathcal B(x_c)=\partial_x\mathcal B(x_c)=0$. Defining
\begin{equation}
    D\equiv m^2-n^2>0,
    \qquad
    q_c\equiv x_c^m,
\end{equation}
one obtains
\begin{equation}
q_c=\frac{D+\sqrt{D(m^2-1)}}{n^2-1}
\label{eq:qc_m_gt_n}
\end{equation}
and therefore
\begin{equation}
\beta_c^n=
\frac{x_c^n\left[(m+n)+(n+1)q_c\right]}
{(n-1)q_c-(m-n)},
\qquad x_c=q_c^{1/m}.
\label{eq:betac_m_gt_n}
\end{equation}
For $\beta<\beta_c$ the deformed family has only the $2T$ sector; at $\beta=\beta_c$ the two noncentral bifurcations coalesce; and for $\beta>\beta_c$ the finite $4T$ window in Eq.~\eqref{eq:four_throat_window} opens. For example, $(m,n)=(6,4)$ gives $\beta_c\simeq1.65081$.

The resulting classification is summarized in Table~\ref{tab:topology_classification}. Away from the bifurcation boundaries, the family therefore supports at most four throats.

\begin{table}[t]
\centering
\caption{Global topology of the profile \eqref{eq:master_function}. The symbols $a_c$ and $a_*$ denote, respectively, the central and noncentral critical amplitudes defined in Eqs.~\eqref{eq:critical_A_equal_mn} and \eqref{eq:bifurcation_amplitude}.}
\label{tab:topology_classification}
\begin{tabular}{c c c}
\hline\hline
Parameter sector & \qquad Central extremum & \qquad Allowed regular topologies \\
\hline
$n>m$ &\quad throat & $1T$ or $3T$ \\
$m=n$, $\beta\leq\beta_c$ & \quad throat/belly across $a_c$ & $1T$ or $2T$ \\
$m=n$, $\beta>\beta_c$ & \quad throat/belly across $a_c$ & $1T$, $3T$, or $2T$ \\
$m>n$, $a=0$ &\quad throat & $1T$ \\
$m>n$, $a>0$ &\quad belly & $2T$ or $4T$ \\
\hline\hline
\end{tabular}
\end{table}

The two-throat sector remains the main configuration used below for the analysis of matter sources and geodesics, but it should be understood as one member of the larger multi-throat family defined by Eq.~\eqref{eq:master_function}.

\subsection{Reference two-throat configuration}
\label{subsec:fiducial_geometry}

To keep the subsequent physical analysis internally consistent, we adopt a single reference configuration whenever a representative two-throat geometry is required. Unless explicitly stated otherwise, we use
\begin{equation}
    m=n=4,\qquad r_0=b=1,\qquad \alpha=0.5,\qquad A=0.25.
    \label{eq:fiducial_parameters}
\end{equation}
All dimensional quantities in this reference configuration are measured in units of $r_0$. In terms of the variables in Eq.~\eqref{eq:dimensionless_variables},
\begin{equation}
    \beta=1,\qquad a=0.5.
\end{equation}
For $m=n=4$ the central transition occurs at
\begin{equation}
    a_c=\frac{\beta^4}{4}=0.25,\qquad
    A_c=0.125,
\end{equation}
while
\begin{equation}
    \beta_c=\left(\frac{8}{3}\right)^{1/4}\simeq1.278.
\end{equation}
Since $\beta<\beta_c$ and $a>a_c$, this point lies unambiguously in the direct $1T\to2T$ branch of the equal-power sector; no intermediate three-throat window is crossed. The center is a degenerate belly: using Eq.~\eqref{eq:central_expansion},
\begin{equation}
    r(\ell)-r(0)=-\frac{1}{8}\ell^4+\mathcal O(\ell^8),
\end{equation}
so the first nonvanishing even derivative confirms a local maximum even though $r''(0)=0$.

The two symmetric throats are located numerically at
\begin{equation}
    \ell_{\rm th}=\pm0.927832,
\end{equation}
with areal radius
\begin{equation}
    r_{\rm th}=0.717936,
\end{equation}
and the central belly has
\begin{equation}
    r_{\rm B}=r(0)=0.75.
\end{equation}
Accordingly, the proper inter-throat separation is
\begin{equation}
    \Delta L=2\ell_{\rm th}\simeq1.855664.
\end{equation}
We use this parameter set as a fixed reference when comparing the matter sector, energy conditions, and geodesic structure. Other choices appear only where the one-, three-, or four-throat sectors need to be illustrated.

\subsection{Embedding criterion}

For a static equatorial slice, $t=\mathrm{const.}$ and $\theta=\pi/2$, the induced metric is
\begin{equation}
    ds^2_{\Sigma}=d\ell^2+r^2(\ell)d\phi^2.
\end{equation}
An isometric embedding into Euclidean three-space with cylindrical metric $ds_E^2=dz^2+dr^2+r^2d\phi^2$ satisfies
\begin{equation}
    \frac{dz}{d\ell}=\pm\sqrt{1-r'(\ell)^2}.
    \label{eq:embedding}
\end{equation}
A global Euclidean embedding of the slice therefore requires
\begin{equation}
    |r'(\ell)|\leq1
\end{equation}
throughout the domain represented by the embedding. We use this condition only as a diagnostic for the visualization of selected configurations; the throat classification itself follows directly from the extrema of $r(\ell)$ and does not depend on the existence of an embedding in Euclidean three-space. For the reference configuration in Eq.~\eqref{eq:fiducial_parameters}, a direct numerical check gives $|r'(\ell)|<1$ throughout the full radial domain, with $|r'|\to\alpha=0.5$ asymptotically. It therefore admits the global Euclidean embedding used in the geometric visualizations below.

\begin{figure*}[t]
    \centering
    \includegraphics[width=\textwidth]{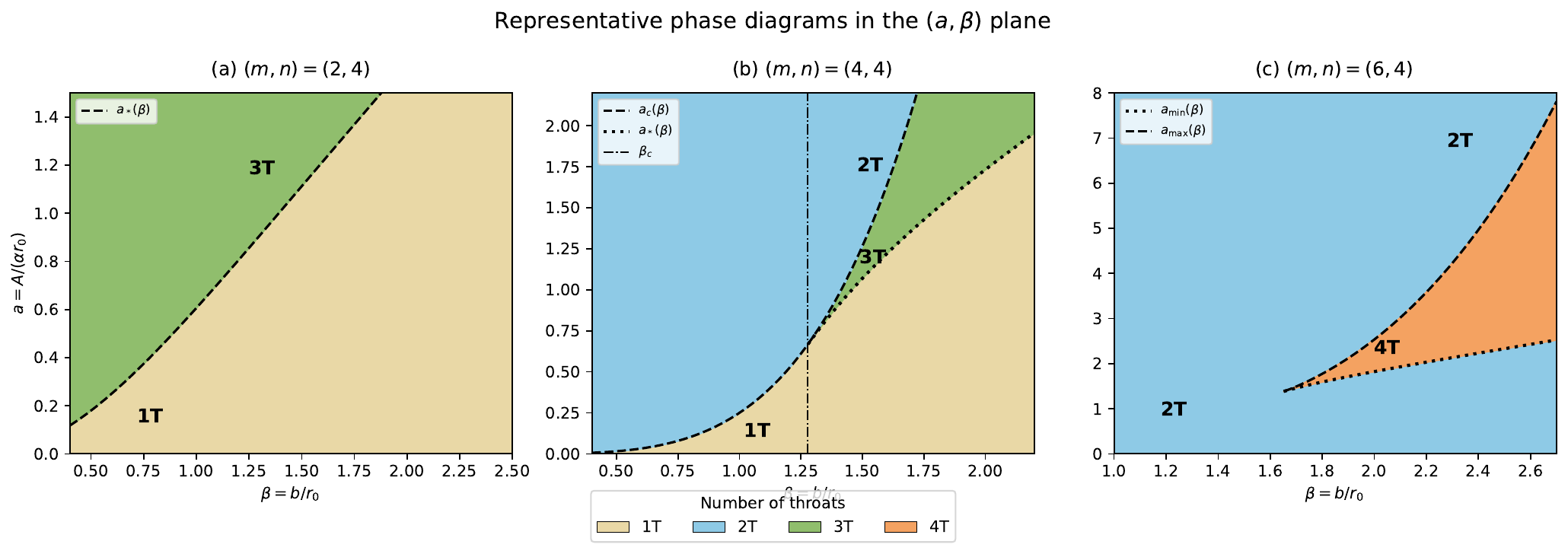}
    \caption{Representative phase diagrams in the $(a,\beta)$ plane. Colors encode the number of throats and are kept consistent across all panels: $1T$ (beige), $2T$ (blue), $3T$ (green), and $4T$ (orange). Panel (a) corresponds to the $n>m$ sector, illustrated here by $(m,n)=(2,4)$, where the bifurcation line $a=a_*(\beta)$ separates the $1T$ and $3T$ regions. Panel (b) corresponds to the equal-power sector $(m,n)=(4,4)$: for $\beta\leq\beta_c$ the critical line $a_c(\beta)$ separates $1T$ from $2T$, whereas for $\beta>\beta_c$ an intermediate $3T$ window opens between $a_*(\beta)$ and $a_c(\beta)$. Panel (c) illustrates the deformed $m>n$ sector with $(m,n)=(6,4)$ and $a>0$, in which the $4T$ window is bounded by $a_{\min}(\beta)$ and $a_{\max}(\beta)$; the undeformed boundary $a=0$, which has zero area in the diagram, corresponds to the original $1T$ generalized Ellis geometry.}
    \label{fig:phase_diagrams}
\end{figure*}

Figure~\ref{fig:phase_diagrams} summarizes the global topology in the $(a,\beta)$ plane for three representative discrete sectors. Panel~(a) illustrates the case $n>m$, for which only the $1T$ and $3T$ sectors occur, separated by the noncentral bifurcation line $a=a_*(\beta)$. Panel~(b) shows the equal-power sector $m=n$, where the direct $1T\to2T$ transition valid for $\beta\leq\beta_c$ is replaced, for $\beta>\beta_c$, by the sequence $1T\to3T\to2T$ bounded by $a_*(\beta)$ and $a_c(\beta)$. Panel~(c) illustrates the sector $m>n$, in which a finite $4T$ window opens between the two noncentral branches $a_{\min}(\beta)$ and $a_{\max}(\beta)$.

The analytic classification and Fig.~\ref{fig:phase_diagrams} already encode the one-dimensional extremum structure. For geometric visualization we therefore use two-dimensional isometric embeddings, with $r(\ell)$ on the horizontal axis and $z(\ell)$ on the vertical axis.

\begin{figure*}[t]
    \centering
    \includegraphics[width=\textwidth]{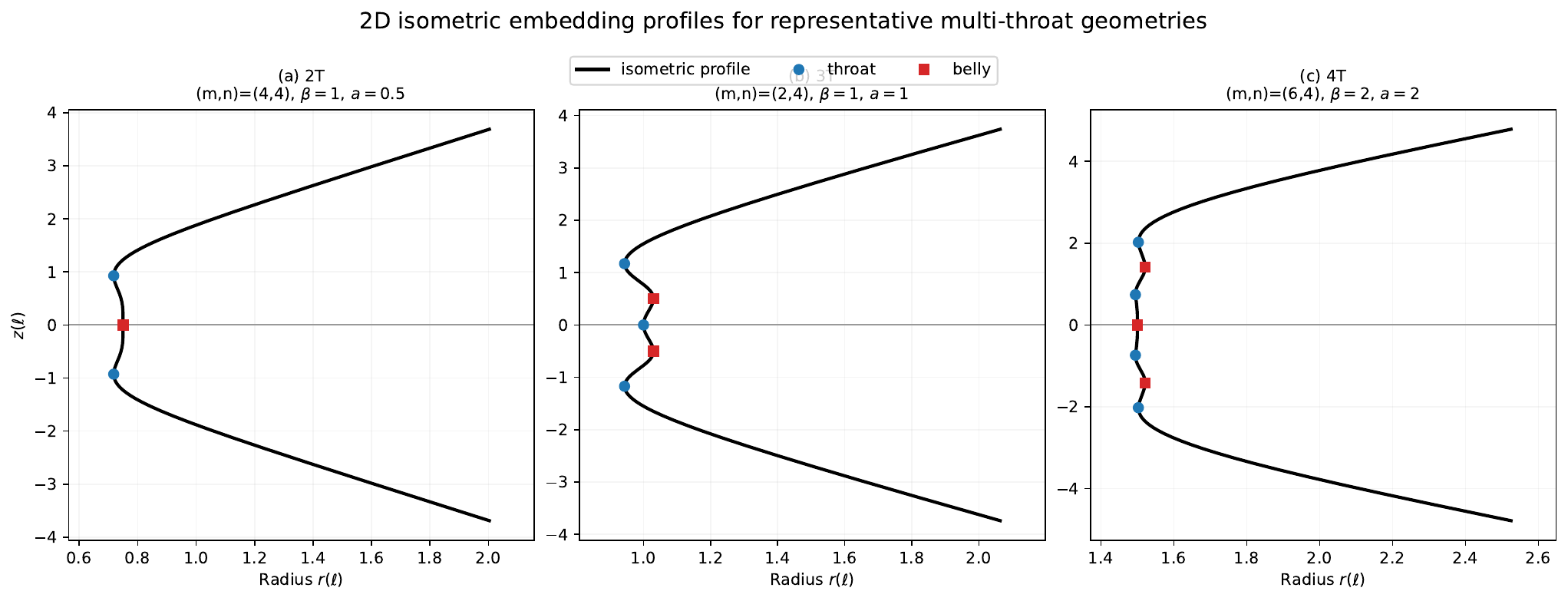}
    \caption{Two-dimensional isometric embedding profiles for representative multi-throat geometries. Panel (a) shows the reference $2T$ configuration with $(m,n)=(4,4)$, $\beta=1$, and $a=0.5$. Panel (b) shows a $3T$ configuration with $(m,n)=(2,4)$, $\beta=1$, and $a=1$. Panel (c) shows a $4T$ configuration with $(m,n)=(6,4)$, $\beta=2$, and $a=2$. In all panels $r_0=1$ and $\alpha=0.5$. Blue circles mark throats and red squares mark bellies.}
    \label{fig:embedding_examples}
\end{figure*}

Figure~\ref{fig:embedding_examples} shows such 2D isometric profiles for three representative configurations that satisfy the global embedding condition $|r'(\ell)|<1$: the reference double-throat configuration, a three-throat example from the $n>m$ sector, and a four-throat example from the $m>n$ sector. In all three cases we use $r_0=1$ and $\alpha=0.5$. Blue circular markers denote throats, while red square markers denote bellies.

\subsection{Regularity, curvature invariants, and asymptotic geometry}
\label{subsec:curvature_regular}

The family in Eq.~\eqref{eq:master_function} is everywhere regular for the parameter range considered here. Since $m$ and $n$ are even, $r_0>0$, $b>0$, $A\geq0$, and $\alpha>0$, one has
\begin{equation}
    r(\ell)>0
\end{equation}
for all $\ell\in\mathbb R$, and the profile is a smooth function over the entire proper-radial domain. Consequently, throats, bellies, the noncentral stationary-inflection folds, and the even-order central degeneracies identified above are not curvature singularities. Moreover, because $\ell$ spans the full real line and the continuous positive function $r(\ell)$ diverges at both ends, it has a strictly positive global minimum. The spatial metric $d\ell^2+r^2(\ell)d\Omega^2$ is therefore complete; the corresponding ultrastatic product spacetime is geodesically complete.

For the ultrastatic metric \eqref{eq:metric}, it is convenient to define
\begin{equation}
    \mathcal A_c(\ell)=\frac{r''}{r},
    \qquad
    \mathcal B_c(\ell)=\frac{1-r'^2}{r^2}.
\end{equation}
The Ricci scalar, the quadratic Ricci invariant, the Kretschmann scalar, and the quadratic Weyl invariant then take the compact forms
\begin{align}
    R&=-4\mathcal A_c+2\mathcal B_c,
    \label{eq:ricci_scalar_compact}\\
    R_{\mu\nu}R^{\mu\nu}
    &=4\mathcal A_c^2+2\left(-\mathcal A_c+\mathcal B_c\right)^2,
    \label{eq:ricci_squared_compact}\\
    K&\equiv R_{\mu\nu\rho\sigma}R^{\mu\nu\rho\sigma}
    =4\left(2\mathcal A_c^2+\mathcal B_c^2\right),
    \label{eq:kretschmann_compact}\\
    C_{\mu\nu\rho\sigma}C^{\mu\nu\rho\sigma}
    &=\frac{4}{3}\left(\mathcal A_c+\mathcal B_c\right)^2.
    \label{eq:weyl_compact}
\end{align}
These expressions make the regularity transparent: because $r(\ell)$ never vanishes and the metric coefficients are smooth and nondegenerate for every finite $\ell$, the orthonormal components of the Riemann tensor remain finite. Hence any algebraic polynomial scalar constructed from the Riemann tensor and the metric is finite throughout the spacetime; the four invariants displayed above provide explicit representatives of this general statement.

A useful point concerns degenerate central extrema. For $m,n>2$ one may have $r'(0)=r''(0)=0$, but this does \emph{not} imply local flatness. At such a point,
\begin{equation}
    R(0)=\frac{2}{r(0)^2},\qquad
    R_{\mu\nu}R^{\mu\nu}(0)=\frac{2}{r(0)^4},
\end{equation}
\begin{equation}
    K(0)=\frac{4}{r(0)^4},\qquad
    C_{\mu\nu\rho\sigma}C^{\mu\nu\rho\sigma}(0)
    =\frac{4}{3r(0)^4}.
\end{equation}
Thus, $r''(0)=0$ only removes the radial contribution $\mathcal A_c$ at the center; the angular curvature $\mathcal B_c=1/r(0)^2$ remains nonzero.

\begin{figure*}[t]
    \centering
    \includegraphics[width=0.95\textwidth]{ 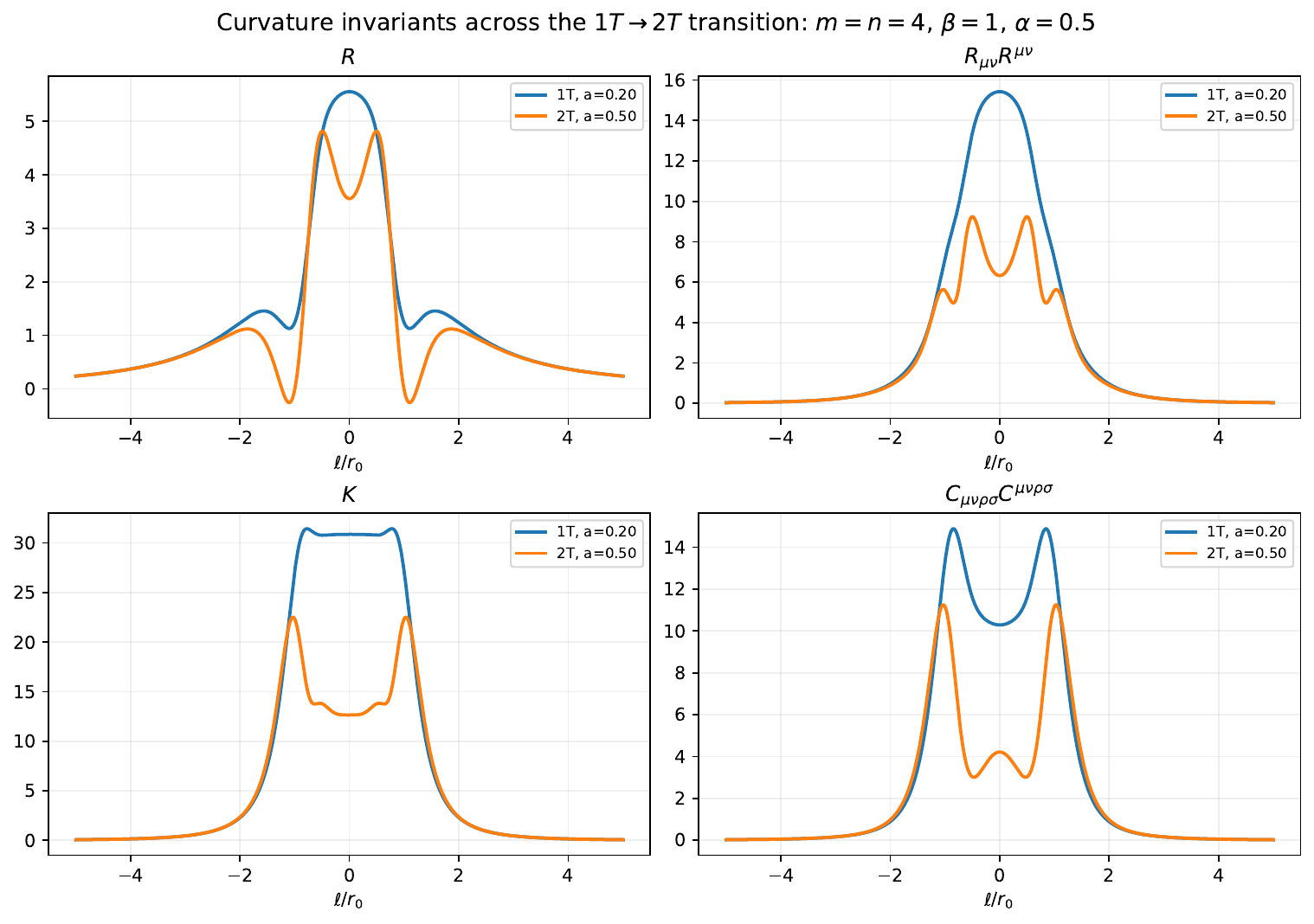}
    \caption{Curvature invariants across the direct $1T\to2T$ bifurcation for $m=n=4$, $\beta=1$, and $\alpha=0.5$. The one-throat configuration has $a=0.20$, while the two-throat curve corresponds to the reference configuration $a=0.50$. The panels show, respectively, the Ricci scalar $R$, the quadratic Ricci invariant $R_{\mu\nu}R^{\mu\nu}$, the Kretschmann scalar $K$, and the quadratic Weyl invariant $C_{\mu\nu\rho\sigma}C^{\mu\nu\rho\sigma}$. The invariants remain finite throughout the transition.}
    \label{fig:curvature_bifurcation}
\end{figure*}

Figure~\ref{fig:curvature_bifurcation} compares the curvature invariants on the two sides of the direct $1T\to2T$ transition in the equal-power sector with $m=n=4$, $\beta=1$, and $\alpha=0.5$. The one-throat curve corresponds to $a=0.20<a_c$, whereas the two-throat curve is the reference configuration $a=0.50>a_c$. All invariants remain finite while their central structure reorganizes across the topological bifurcation.

The asymptotic structure follows directly from Eq.~\eqref{eq:master_function}. For $|\ell|\to\infty$,
\begin{align}
    r(\ell)=&\,\alpha|\ell|
    +\frac{\alpha r_0^m}{m}|\ell|^{1-m}
    +A b^n|\ell|^{-n}+\cdots,
    \label{eq:r_asymptotic}\\
    r'(\ell)=&\,\alpha\,\mathrm{sgn}(\ell)
    +\mathcal O(|\ell|^{-m})+\mathcal O(|\ell|^{-n-1}),\\
    r''(\ell)=&\,\alpha(m-1)r_0^m|\ell|^{-m-1}
    +n(n+1)A b^n|\ell|^{-n-2}+\cdots.
    \label{eq:rpp_asymptotic}
\end{align}
Hence the curvature invariants vanish asymptotically according to
\begin{align}
    R&\sim \frac{2(1-\alpha^2)}{r^2},\\
    R_{\mu\nu}R^{\mu\nu}&\sim
    \frac{2(1-\alpha^2)^2}{r^4},\\
    K&\sim\frac{4(1-\alpha^2)^2}{r^4},\\
    C_{\mu\nu\rho\sigma}C^{\mu\nu\rho\sigma}
    &\sim\frac{4(1-\alpha^2)^2}{3r^4}.
\end{align}
Although all curvature invariants tend to zero, the spacetime is not globally asymptotically Minkowskian when $\alpha<1$. Introducing the asymptotic areal radius $R=\alpha|\ell|$, the metric approaches
\begin{equation}
    ds^2\simeq-dt^2+\frac{dR^2}{\alpha^2}+R^2d\Omega^2,
    \label{eq:asymptotic_conical_metric}
\end{equation}
which is a conical geometry with solid-angle deficit
\begin{equation}
    \Delta\Omega=4\pi(1-\alpha^2).
\end{equation}
The corresponding stress-energy interpretation is developed in Sec.~\ref{subsec:string_cloud}.

\section{Field Sources and the Asymptotic String-Cloud Sector}
\label{sec:field_sources}

A one-field reconstruction with a sign-changing kinetic prefactor can reproduce the background geometry, but such a parametrization becomes kinetically degenerate at the canonical--phantom transition surfaces. This is the same structural issue encountered in trapped-ghost constructions \cite{BronnikovSushkov2010,BronnikovReview2018}. We therefore adopt a two-scalar realization with fixed kinetic signs, following the general strategy of canonical--phantom source models such as Ref.~\cite{BronnikovKorolyov2015}. This removes the zero crossing of a field-dependent kinetic coefficient, but it does not eliminate the phantom degree of freedom itself.

\subsection{Two-scalar reconstruction with a magnetic nonlinear electromagnetic sector}

We consider the action
\begin{equation}
S=\int d^4x\sqrt{-g}
\left[
\frac{R}{2\kappa}
-\frac12(\partial\chi)^2
+\frac12(\partial\psi)^2
-U(\chi)
-f(\chi)\,\mathcal G(F)
\right],
\label{eq:two_scalar_action}
\end{equation}
where $\chi$ is canonical and $\psi$ is phantom. We define the electromagnetic invariant by
\begin{equation}
F\equiv F_{\mu\nu}F^{\mu\nu},
\end{equation}
and use the magnetic monopole ansatz
\begin{equation}
F_{\theta\phi}=P\sin\theta,
\qquad
F=\frac{2P^2}{r^4(\ell)}.
\label{eq:F_magnetic_source}
\end{equation}
The nonlinear-electromagnetic model is chosen as
\begin{equation}
\mathcal G(F)=F+\lambda\sqrt F,
\qquad \lambda>0,
\label{eq:Gmodel_twoscalar}
\end{equation}
for which
\begin{equation}
\mathcal G_F=1+\frac{\lambda}{2\sqrt F},
\qquad
\mathcal G_F+2F\mathcal G_{FF}=1.
\label{eq:Gcharacteristics_twoscalar}
\end{equation}
The Maxwell term therefore lifts the degeneracy of the second combination, which would vanish for a pure $\sqrt F$ model. We take the real branch $\sqrt F$, so the model is defined for $F\geq0$; this covers the purely magnetic background, for which $F=2P^2/r^4>0$. Electric-dominated configurations or perturbations with $F<0$ fall outside this effective model. Since the square-root term dominates as $F\to0$, the weak-field limit is non-Maxwellian. The same $\sqrt F\sim r^{-2}$ scaling assigns the leading string-cloud-like stress to the NED sector.

For radial scalar fields, the matter stress tensor generated by Eq.~\eqref{eq:two_scalar_action} can be written as
\begin{align}
\rho&=\frac12\left(\chi'^2-\psi'^2\right)+U+f\mathcal G,
\label{eq:rho_twoscalar_matter}\\
p_r&=\frac12\left(\chi'^2-\psi'^2\right)-U-f\mathcal G,
\label{eq:pr_twoscalar_matter}\\
p_t&=-\frac12\left(\chi'^2-\psi'^2\right)-U-f\mathcal G
+2fF\mathcal G_F.
\label{eq:pt_twoscalar_matter}
\end{align}
Hence
\begin{equation}
\rho+p_r=\chi'^2-\psi'^2,
\qquad
\rho+p_t=2fF\mathcal G_F.
\label{eq:null_combinations_twoscalar}
\end{equation}
These relations provide the direct bridge between the prescribed geometry and the reconstructed matter functions. In particular, for radial scalars the scalar kinetic and potential terms cancel identically from $\rho+p_t$; within the present ansatz this combination is carried entirely by the scalar-coupled NED sector.

We choose the potential and the electromagnetic coupling to depend on the canonical scalar $\chi$, while the phantom field $\psi$ remains purely kinetic and shift symmetric. This keeps the phantom equation analytically integrable and makes $\chi$ a monotonic field-space coordinate along the solution, so both $U$ and $f$ can be reconstructed as single-valued functions of it. Other assignments, including functions $U(\chi,\psi)$ and $f(\chi,\psi)$, are possible. The background geometry fixes such functions only along the one-dimensional trajectory $(\chi(\ell),\psi(\ell))$, leaving their off-trajectory extension nonunique.

Scalar--electromagnetic source reconstructions have been used previously for generalized and multi-throat wormholes \cite{CrispimMulti2025,CrispimGEB2026,SuHao2025}. In the present construction there is, however, a stronger reason for retaining a scalar-dependent coupling in front of the NED sector. If one tried to set $f=1$, Eq.~\eqref{eq:null_combinations_twoscalar} together with the Einstein equations would require
\begin{equation}
\mathcal G_F(F)=-\frac{r^2Q}{4\kappa P^2},
\label{eq:GF_uncoupled_obstruction}
\end{equation}
where
\begin{equation}
Q(\ell)\equiv r'^2-1+r r''.
\label{eq:Qdef_twoscalar}
\end{equation}
For the magnetic monopole, $F=2P^2/r^4$ depends only on the areal radius. If $r(\ell)$ is nonmonotonic and $Q$ is not single-valued as a function of $r$, two radial positions can share the same $F$ while Eq.~\eqref{eq:GF_uncoupled_obstruction} demands different values of $\mathcal G_F$. The reference double-throat configuration provides an explicit example: $r=0.74$ occurs at $\ell\simeq0.552$ and $\ell\simeq1.182$, where $Q\simeq-1.145$ and $Q\simeq-0.464$. In units $\kappa=P=1$, the corresponding values required for $\mathcal G_F$ are approximately $0.1567$ and $0.06346$. A single-valued uncoupled $\mathcal G(F)$ cannot satisfy both conditions.

This obstruction follows under the assumptions used here: a purely magnetic monopole, radial scalar fields, an uncoupled electromagnetic sector when $f=1$, and a single-valued Lagrangian depending only on $F$. Multi-branch NED descriptions are a known alternative in regular electric constructions \cite{Bronnikov2001NED,BronnikovBranch2017}. Here we keep a single-valued electromagnetic theory and use $f(\chi)$ to distinguish points that have the same $F$. The monotonic phantom field $\psi$ could play the same labeling role in a different reconstruction; assigning the coupling to $\chi$ is convenient because the potential already depends on that field. The scalar coupling is therefore a branch-resolving device for the chosen matter ansatz, not a uniqueness statement about multi-throat sources.

We choose the phantom field to be shift-symmetric, so that its equation of motion reduces to
\begin{equation}
(r^2\psi')'=0.
\end{equation}
Hence
\begin{equation}
\psi'=\frac{q}{r^2},
\label{eq:psi_prime}
\end{equation}
where $q$ is a constant scalar charge. Combining Eq.~\eqref{eq:null_combinations_twoscalar} with the geometric radial-null relation gives
\begin{equation}
\kappa(\rho+p_r)=-\frac{2r''}{r}
=\kappa\left(\chi'^2-\psi'^2\right),
\label{eq:chi_psi_nec_r}
\end{equation}
and therefore
\begin{equation}
\chi'^2=\frac{q^2}{r^4}-\frac{2r''}{\kappa r}.
\label{eq:chi_prime_sq}
\end{equation}
This relation reproduces the required change in the radial NEC without any zero crossing of a kinetic coefficient. Where $r''>0$, the phantom gradient dominates; where $r''<0$, the canonical gradient dominates; and at surfaces with $r''=0$ their kinetic densities are equal.

Reality of $\chi$ requires
\begin{equation}
q^2\ge q_{\rm crit}^2,
\qquad
q_{\rm crit}^2\equiv \frac{2}{\kappa}
\max_{\ell}\bigl[r^3(\ell)r''(\ell)\bigr].
\label{eq:qcrit}
\end{equation}
For the reference two-throat configuration of Eq.~\eqref{eq:fiducial_parameters}, in the normalized source plots we set $r_0=\kappa=1$, and a direct numerical maximization gives
\begin{equation}
q_{\rm crit}\simeq0.74814.
\end{equation}
If instead one uses the later convention $G=1$, so that $\kappa=8\pi$, the same geometry gives $q_{\rm crit}\simeq0.14923$. The representative choice
\begin{equation}
q=1>q_{\rm crit}
\label{eq:q_choice}
\end{equation}
lies safely above the threshold in either convention.
Choosing the positive branch of $\chi'$, both $\chi'(\ell)$ and $\psi'(\ell)$ are strictly positive at every finite $\ell$. Thus both scalar fields are smooth and monotonic and can be inverted globally on their respective images.

The tangential null combination is
\begin{equation}
\kappa(\rho+p_t)=-\frac{Q}{r^2}
=2\kappa fF\mathcal G_F.
\label{eq:chi_psi_nec_t}
\end{equation}
It follows that
\begin{equation}
f(\chi)=
-\left.
\frac{r^2Q}{4\kappa P^2\mathcal G_F}
\right|_{\ell=\ell(\chi)},
\label{eq:fchi_general}
\end{equation}
and, using $\rho-p_r=2(U+f\mathcal G)$ together with the Einstein equations,
\begin{equation}
U(\chi)=
-\left.\frac{Q}{\kappa r^2}\right|_{\ell=\ell(\chi)}
-f(\chi)\,\mathcal G(F).
\label{eq:Uchi_general}
\end{equation}
With $s\equiv\sqrt F$, these become
\begin{equation}
f(\chi)=
-\left.
\frac{r^2Q}{4\kappa P^2}
\frac{1}{1+\lambda/(2s)}
\right|_{\ell=\ell(\chi)},
\label{eq:fchi_regularized}
\end{equation}
\begin{equation}
U(\chi)=
-\left.
\frac{Q}{\kappa r^2}
\frac{s}{2s+\lambda}
\right|_{\ell=\ell(\chi)}.
\label{eq:Uchi_regularized}
\end{equation}
For the reference configuration, a direct global check gives $Q(\ell)<0$, with $Q\to\alpha^2-1=-0.75$ asymptotically. Hence $f(\chi)>0$. Writing the electromagnetic Lagrangian as $\mathcal L(F,\chi)=-f(\chi)\mathcal G(F)$, one has
\begin{equation}
\mathcal L_F=-f\mathcal G_F
=\frac{r^2Q}{4\kappa P^2}<0,
\end{equation}
and
\begin{equation}
\mathcal L_F+2F\mathcal L_{FF}
=-f\left(\mathcal G_F+2F\mathcal G_{FF}\right)
=-f<0.
\end{equation}
These two derivative combinations therefore have the Maxwell sign throughout the reference domain. The sign check is a background diagnostic only; hyperbolicity and perturbative stability require the coupled characteristic problem. Because $f_\chi\neq0$, scalar and electromagnetic perturbations can mix on the magnetic background.

\subsection{Consistency of the reconstructed matter equations}

Because the matter functions were obtained by inverse reconstruction from the Einstein equations, it is important to verify that the independent matter equations are also satisfied. Variation of Eq.~\eqref{eq:two_scalar_action} gives
\begin{align}
\Box\psi&=0,
\label{eq:psi_eom_full}\\
\Box\chi-U_{\chi}-f_{\chi}\mathcal G&=0,
\label{eq:chi_eom_full}\\
\nabla_{\mu}\left(f\mathcal G_F F^{\mu\nu}\right)&=0.
\label{eq:NED_eom_full}
\end{align}
Equation~\eqref{eq:psi_eom_full} is precisely the equation integrated in Eq.~\eqref{eq:psi_prime}. For the purely magnetic monopole $F_{\theta\phi}=P\sin\theta$, Eq.~\eqref{eq:NED_eom_full} is satisfied identically: the only nonzero contravariant electromagnetic component is angular, and $\sqrt{-g}\,f\mathcal G_F F^{\theta\phi}=P f\mathcal G_F/r^2$ is independent of the angular coordinates on which the divergence acts.

The canonical-scalar equation follows from the remaining background equations. Once the Einstein equations, Eq.~\eqref{eq:psi_eom_full}, and Eq.~\eqref{eq:NED_eom_full} hold, the contracted Bianchi identity gives
\begin{equation}
\chi'\left(
\chi''+2\frac{r'}{r}\chi'
-U_{\chi}-f_{\chi}\mathcal G
\right)=0.
\end{equation}
For the choice $q>q_{\rm crit}$ adopted above, $\chi'$ does not vanish at any finite radial position, so the bracket is identically zero and Eq.~\eqref{eq:chi_eom_full} follows. The inverse reconstruction therefore closes the complete background field equations, not only the Einstein sector.

\begin{figure*}[t]
    \centering
    \includegraphics[width=\textwidth]{ 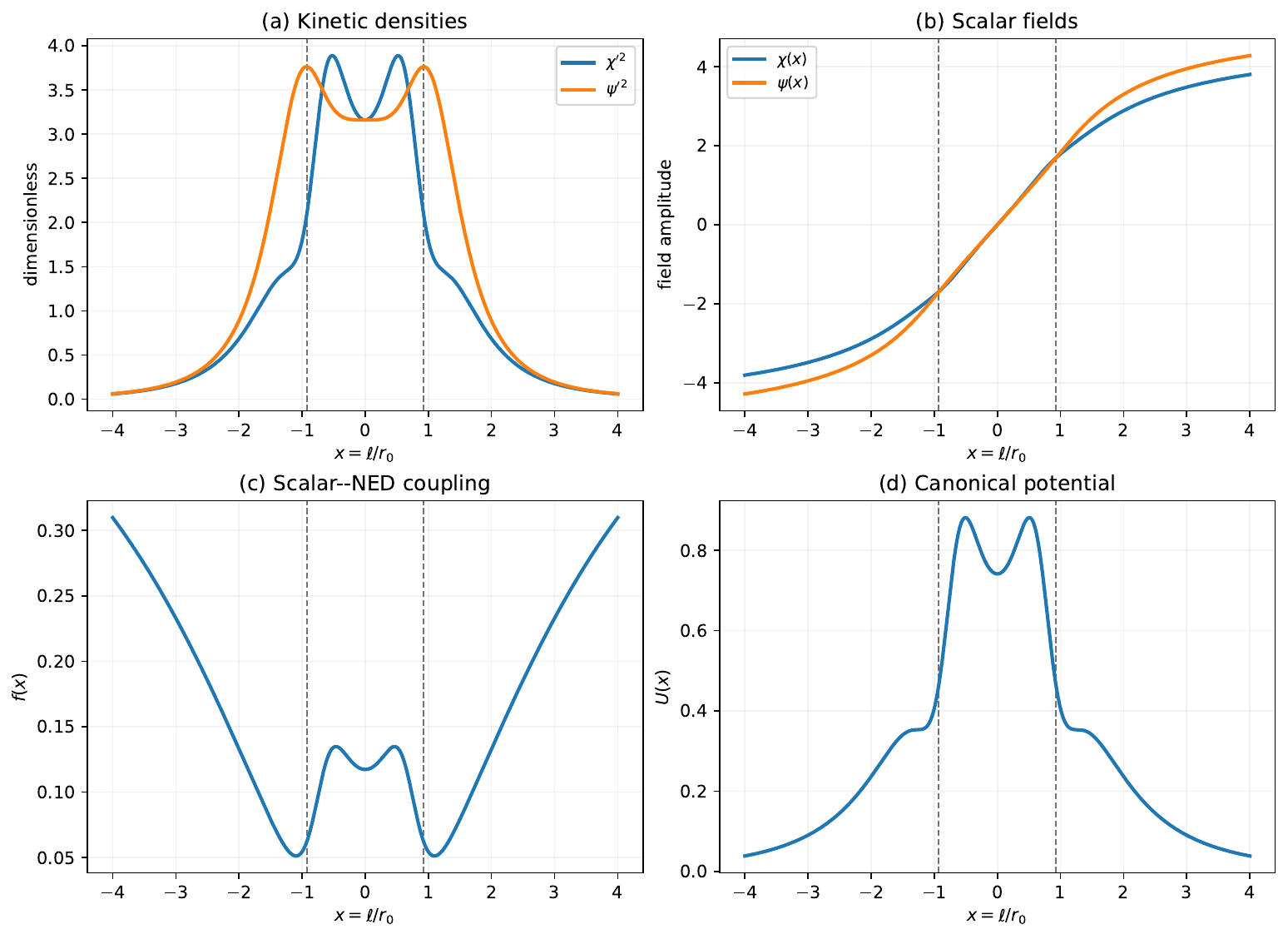}
    \caption{Reconstructed source functions for the reference two-throat geometry in the two-scalar model, using $q=1$ and normalized plotting units $r_0=\kappa=P=\lambda=1$. Panel (a) compares the canonical and phantom kinetic densities, $\chi'^2$ and $\psi'^2$. Panel (b) shows the monotonic scalar fields $\chi(x)$ and $\psi(x)$, with $x\equiv\ell/r_0$. Panel (c) displays the scalar--NED coupling $f(x)$, and panel (d) the canonical potential $U(x)$. The dashed vertical lines mark the two throats at $|x|\simeq0.928$.}
    \label{fig:source_functions_fiducial}
\end{figure*}

Figure~\ref{fig:source_functions_fiducial} displays the reconstructed source functions for the reference configuration. Panel~(a) shows that $\chi'^2\geq\psi'^2$ in the central inter-throat sector, with equality at the degenerate center and at the two $r''=0$ transition surfaces, whereas $\psi'^2>\chi'^2$ in the throat and exterior sectors. Panel~(b) confirms that both scalar fields are smooth and monotonic. Panels~(c) and (d) show the reconstructed scalar--NED coupling and canonical potential.

\subsection{Asymptotic string-cloud behavior}\label{subsec:string_cloud}

Using the large-$|\ell|$ expansion derived in Sec.~\ref{subsec:curvature_regular}, one has
\begin{equation}
Q\longrightarrow\alpha^2-1<0,
\qquad
F\sim\frac{2P^2}{r^4},
\qquad
s=\sqrt F\sim r^{-2}.
\end{equation}
Equations~\eqref{eq:psi_prime}, \eqref{eq:chi_prime_sq}, and \eqref{eq:Uchi_regularized} then imply
\begin{equation}
\psi'^2=\mathcal O(r^{-4}),
\qquad
\chi'^2=\mathcal O(r^{-4}),
\qquad
U=\mathcal O(r^{-4}).
\end{equation}
For $P\neq0$, Eq.~\eqref{eq:fchi_regularized} also gives the finite asymptotic coupling
\begin{equation}
f\longrightarrow
\frac{1-\alpha^2}{\sqrt{2}\,\kappa |P|\lambda}.
\label{eq:f_asymptotic}
\end{equation}
Since the $\sqrt F$ term dominates $\mathcal G(F)$ at small $F$, the nonlinear-electromagnetic contribution behaves as
\begin{equation}
-f\mathcal G(F)\sim\frac{\alpha^2-1}{\kappa r^2}.
\end{equation}
Consequently,
\begin{equation}
\rho\sim\frac{1-\alpha^2}{\kappa r^2},
\qquad
p_r\sim-\rho,
\qquad
p_t\to0,
\label{eq:string_cloud_asymptotic_twoscalar}
\end{equation}
which is the characteristic radial equation of state of an asymptotic string-cloud-like sector \cite{Letelier1979}. With $\mathcal G=F+\lambda\sqrt F$, the $\sqrt F$ term places the leading $r^{-2}$ stress in the electromagnetic sector, while the scalars and $U$ start at order $r^{-4}$. The geometry fixes the total leading stress but not its decomposition among reconstructed fields; a different NED model could redistribute it. The associated angular deficit is closely analogous to the far-field geometry of a global monopole \cite{BarriolaVilenkin1989}.

The inverse reconstruction is nonunique: changing the geometric parameters, the scalar charge $q$, or the NED model generally changes the reconstructed pair $(f,U)$. The present choice avoids the kinetic zero crossing of a one-field sign-changing parametrization but retains a genuine phantom degree of freedom. Simpler ghost-scalar wormholes are radially unstable \cite{GonzalezGuzmanSarbach2009}; the additional scalar--electromagnetic couplings here change the perturbation problem, so their result cannot be imported directly. Establishing dynamical viability requires the coupled metric--scalar--electromagnetic perturbation equations and is left for future work.

\section{Energy Conditions and Exoticity}
\label{sec:energy_conditions}

\subsection{Pointwise energy conditions}

For the ultrastatic metric, the Einstein equations give the compact combinations
\begin{align}
\kappa(\rho+p_r)&=-\frac{2r''}{r},
\label{eq:NEC_radial_analytic}\\
\kappa(\rho+p_t)&=\frac{1-r'^2-r r''}{r^2}
=-\frac{Q}{r^2},
\label{eq:NEC_tangential_analytic}
\end{align}
where $Q=r'^2-1+r r''$ was defined in Sec.~\ref{sec:field_sources}. The radial NEC is therefore violated precisely where $r''>0$, whereas the tangential NEC is satisfied wherever $Q<0$.

The remaining pointwise conditions are
\begin{align}
\text{WEC}:&\qquad \rho\ge0,\quad \rho+p_r\ge0,\quad \rho+p_t\ge0,\\
\text{DEC}:&\qquad \rho\ge |p_r|,\quad \rho\ge |p_t|,\\
\text{SEC}:&\qquad \rho+p_r\ge0,\quad \rho+p_t\ge0,\quad
\rho+p_r+2p_t\ge0.
\end{align}
For every ultrastatic geometry in the present family one has the identity
\begin{equation}
\rho+p_r+2p_t=0,
\label{eq:SEC_saturation}
\end{equation}
so the trace combination entering the SEC is identically saturated. This is the stress-tensor form of $R_{tt}=0$ for an ultrastatic metric. The SEC therefore adds no independent condition: its validity is fixed by the two NEC combinations.

A useful general lemma follows immediately. In any region where
\begin{equation}
r''\leq0,\qquad |r'|<1,
\label{eq:energy_condition_lemma}
\end{equation}
one has $p_r<0$, $p_t\leq0$, $\rho+p_r\geq0$, and $\rho+p_t>0$. These inequalities imply $\rho\geq|p_r|$ and $\rho\geq|p_t|$, so the WEC and DEC hold; the SEC follows from Eq.~\eqref{eq:SEC_saturation}. Related multi-throat reconstructions also contain regions with partially restored pointwise energy conditions \cite{CrispimMulti2025}; Eq.~\eqref{eq:energy_condition_lemma} gives a simple sufficient criterion for the present ultrastatic family. The statement concerns the total stress tensor. The phantom field remains active, $\psi'^2=q^2/r^4>0$, and is locally compensated by the other matter sectors.

We first consider the reference two-throat configuration
\begin{equation}
(m,n)=(4,4),\qquad \beta=1,\qquad \alpha=0.5,\qquad a=0.5,
\end{equation}
which corresponds to $r_0=b=1$ and $A=0.25$. The noncentral zeros of $r''$ occur at
\begin{equation}
|\ell|=\ell_{\rm tr}\simeq0.682252\,r_0,
\label{eq:ell_transition_energy}
\end{equation}
whereas the throats lie at
\begin{equation}
|\ell|=\ell_{\rm th}\simeq0.927832\,r_0.
\end{equation}
For this solution $Q<0$ and $|r'|<1$ globally. Hence, in the open central interval $0<|\ell|<\ell_{\rm tr}$ one has $r''<0$, $p_r<0$, and $p_t<0$. It follows that
\begin{equation}
\rho-|p_r|=\rho+p_r>0,
\qquad
\rho-|p_t|=\rho+p_t>0,
\end{equation}
so the DEC is satisfied there; the WEC follows immediately and the SEC is satisfied because Eq.~\eqref{eq:SEC_saturation} is saturated. At the quartically degenerate central belly, $r'=r''=0$, and therefore
\begin{equation}
\kappa\rho(0)=\frac{1}{r(0)^2},\qquad
p_r(0)=-\rho(0),\qquad
p_t(0)=0.
\label{eq:center_stringlike}
\end{equation}
The radial NEC is saturated at the center and at the transition surfaces and is violated throughout the throat and exterior sectors. The energy density becomes negative only in the narrow intervals
\begin{equation}
0.997141\lesssim |\ell|/r_0\lesssim1.236087,
\end{equation}
before recovering the positive string-cloud asymptotic behavior.

To verify that this pointwise structure is not peculiar to the equal-power two-throat case, we also analyze the representative four-throat configuration used in the geometric survey,
\begin{equation}
(m,n)=(6,4),\qquad \beta=2,\qquad \alpha=0.5,\qquad a=2,
\label{eq:energy_4T_reference}
\end{equation}
corresponding, for $r_0=1$, to $b=2$ and $A=1$. On each positive-$\ell$ branch the extrema occur at
\begin{equation}
\ell_{\rm th}^{(1)}\simeq0.739296,\qquad
\ell_{\rm B}\simeq1.416836,\qquad
\ell_{\rm th}^{(2)}\simeq2.023195,
\end{equation}
with the mirror extrema at negative $\ell$ and a degenerate central belly at $\ell=0$. The positive zeros of $r''$ are
\begin{equation}
\ell_{\rm tr}^{(1)}\simeq0.554612,\qquad
\ell_{\rm tr}^{(2)}\simeq1.084461,\qquad
\ell_{\rm tr}^{(3)}\simeq1.720493.
\label{eq:4T_rpp_zeros}
\end{equation}
A direct check gives $Q_{\max}\simeq-0.416512$ and $|r'|<1$ globally. In fact, $|r'|$ approaches $\alpha=0.5$ asymptotically, so its supremum is $0.5$ rather than the finite-window value seen in a numerical plot. The tangential NEC is therefore positive globally and $p_r<0$ everywhere. By the lemma in Eq.~\eqref{eq:energy_condition_lemma}, the WEC, DEC, and SEC are simultaneously satisfied in the $r''<0$ sectors,
\begin{equation}
|\ell|<0.554612,
\qquad
1.084461<|\ell|<1.720493,
\label{eq:4T_energy_satisfying_bands}
\end{equation}
with the radial NEC saturated at the boundaries and at the degenerate center. The second interval is the finite band surrounding each lateral belly. Conversely, the inner- and outer-throat sectors have $r''>0$ and violate the radial NEC. The energy density becomes negative only beyond the outer throats, in
\begin{equation}
2.115395\lesssim |\ell|\lesssim2.760019.
\end{equation}
Thus the alternating throat--belly structure of the four-throat geometry is mirrored by alternating radial-NEC-violating and radial-NEC-satisfying sectors.

\begin{figure*}[t]
    \centering
    \includegraphics[width=\textwidth]{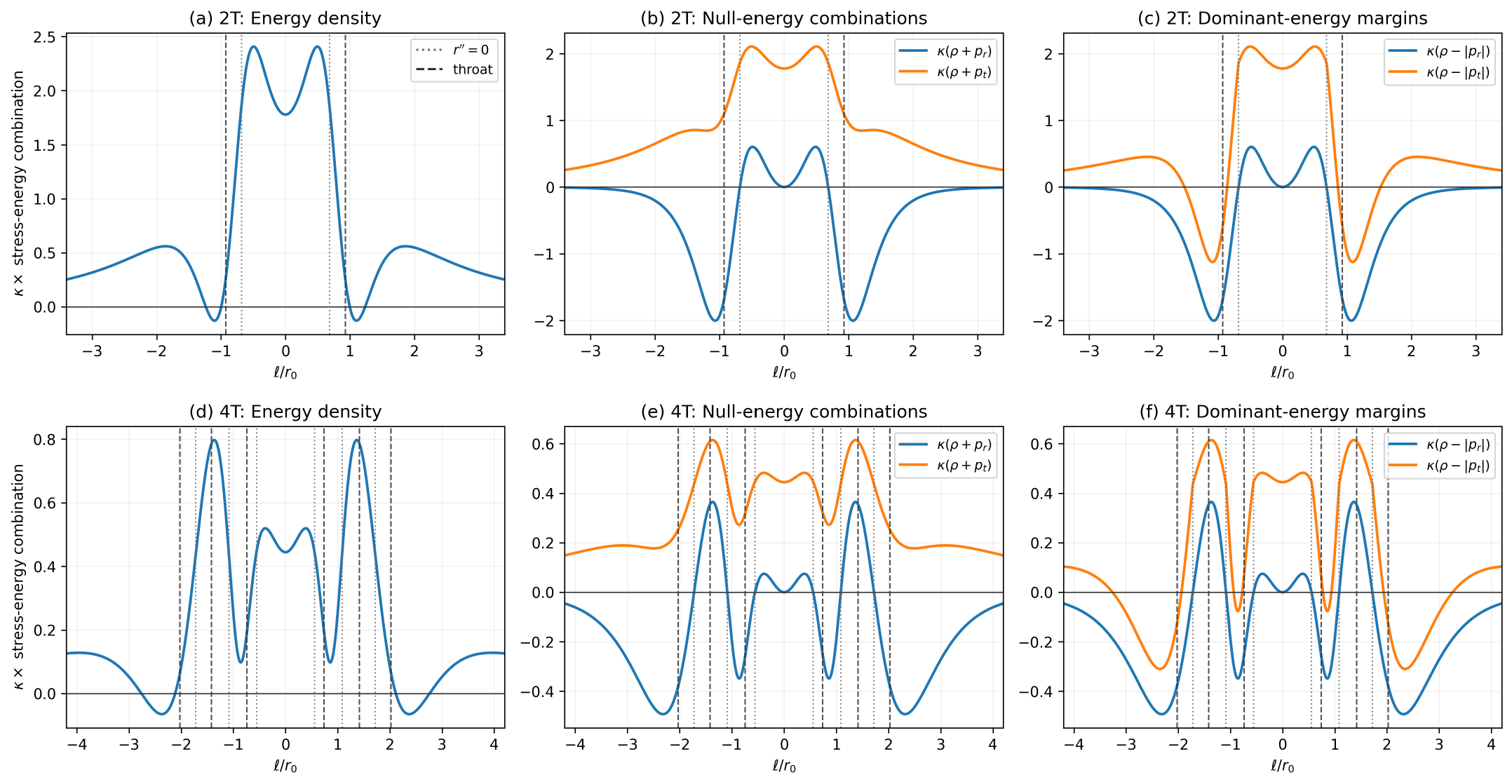}
    \caption{Pointwise energy conditions for representative $2T$ and $4T$ geometries. The upper row corresponds to $(m,n)=(4,4)$, $\beta=1$, $\alpha=0.5$, $a=0.5$; the lower row to $(m,n)=(6,4)$, $\beta=2$, $\alpha=0.5$, $a=2$. Panels (a,d) show $\kappa\rho$; panels (b,e) show $\kappa(\rho+p_r)$ and $\kappa(\rho+p_t)$; panels (c,f) show the dominant-energy margins $\kappa(\rho-|p_r|)$ and $\kappa(\rho-|p_t|)$. Dotted vertical lines mark the zeros of $r''$, where the radial NEC changes sign, and dashed vertical lines mark the throats. In both examples $Q<0$ globally, so the tangential NEC is positive throughout. In the $4T$ case the additional belly produces a second finite band in which the WEC, DEC, and SEC are simultaneously satisfied.}
    \label{fig:energy_conditions_comparison}
\end{figure*}

Figure~\ref{fig:energy_conditions_comparison} compares the two representative geometries. The first column shows $\kappa\rho$, the second the two NEC combinations, and the third the two DEC margins. The SEC is not plotted separately because Eq.~\eqref{eq:SEC_saturation} makes its third combination identically zero; its satisfaction is completely determined by the two NEC curves.

\subsection{Localization and integrability of the NEC-violating sector}

The radial NEC violation is concentrated in amplitude around the throat regions, but it is not compactly supported. Indeed, for $|\ell|\to\infty$,
\begin{equation}
r''(\ell)=
\alpha(m-1)r_0^m|\ell|^{-m-1}
+A n(n+1)b^n|\ell|^{-n-2}+\cdots>0,
\label{eq:rpp_tail_exoticity}
\end{equation}
so that $\rho+p_r$ approaches zero from below. The corresponding proper-volume-weighted contribution is nevertheless integrable, since
\begin{equation}
r r''=\mathcal O(|\ell|^{-m})+\mathcal O(|\ell|^{-n-1}),
\end{equation}
with even $m,n\ge2$. Thus the radial NEC violation has rapidly decaying integrable tails even though it extends to both asymptotic regions.

The two-scalar realization obtained in Sec.~\ref{sec:field_sources} makes this geometric separation transparent:
\begin{equation}
\rho+p_r=\chi'^2-\psi'^2.
\end{equation}
The canonical gradient dominates wherever $r''<0$, while the phantom gradient dominates wherever $r''>0$. The transition occurs through the regular equality $\chi'^2=\psi'^2$, with no vanishing kinetic coefficient.

\subsection{Global Nandi--Zhang--Kumar volume integral and parameter-space structure}

Two closely related volume diagnostics should be distinguished. The Visser--Kar--Dadhich construction uses the areal coordinate and a measure proportional to $4\pi r^2dr$ \cite{VisserEC}. In a multi-throat geometry $r(\ell)$ is not globally one-to-one, so that expression requires a branch decomposition; an unsigned branchwise measure would introduce a factor $|r'|$ when rewritten in $\ell$. Nandi, Zhang, and Kumar instead proposed the covariant measure $\sqrt{-g}\,d^3x$ \cite{NandiEC}. For the present ultrastatic proper-radial metric, $\sqrt{-g}=r^2\sin\theta$, so the Nandi--Zhang--Kumar measure is exactly
\begin{equation}
dV_{\rm NZK}=4\pi r^2d\ell.
\end{equation}
We therefore denote by $I_{\rm VIQ}$ the global Nandi--Zhang--Kumar integral
\begin{equation}
I_{\rm VIQ}
\equiv
\int_{-\infty}^{+\infty}(\rho+p_r)\,dV_{\rm NZK}
=-\frac{8\pi}{\kappa}
\int_{-\infty}^{+\infty}r r''\,d\ell.
\label{eq:VIQ_global}
\end{equation}
Because this is a signed global integral, NEC-satisfying and NEC-violating regions contribute with opposite signs. A positive value therefore does not remove the local radial-NEC violation at the throats, and $I_{\rm VIQ}=0$ means that the positive and negative contributions cancel globally.

For the two reference geometries used above, with $r_0=1$ and $\kappa=8\pi$, numerical integration gives
\begin{equation}
I_{\rm VIQ}^{(2T)}\simeq-0.83435,
\qquad
I_{\rm VIQ}^{(4T)}\simeq-1.72621.
\label{eq:VIQ_reference_values}
\end{equation}
Both examples therefore have negative global VIQ, despite containing finite regions in which all standard pointwise energy conditions are satisfied.

The sign change of the global VIQ can be characterized analytically. Using the dimensionless variables of Sec.~\ref{field_eqs}, write
\begin{equation}
\mathcal R(x)=g_m(x)+a\,h_{n,\beta}(x),
\qquad
g_m(x)=(1+x^m)^{1/m},
\qquad
h_{n,\beta}(x)=\frac{1}{1+(x/\beta)^n}.
\end{equation}
Since $r=\alpha r_0\mathcal R$ and $x=\ell/r_0$, Eq.~\eqref{eq:VIQ_global} becomes
\begin{equation}
I_{\rm VIQ}
=-\frac{16\pi\alpha^2 r_0}{\kappa}
\int_0^\infty \mathcal R\mathcal R''\,dx.
\label{eq:IVIQ_dimensionless}
\end{equation}
Integrating the mixed and pure-deformation terms by parts gives
\begin{equation}
I_{\rm VIQ}
=\frac{16\pi\alpha^2 r_0}{\kappa}
\left(K_2a^2-J_1a-J_0\right),
\label{eq:IVIQ_quadratic}
\end{equation}
where
\begin{align}
J_0&\equiv\int_0^\infty g_m g_m''\,dx>0,\\
J_1&\equiv-2\int_0^\infty g_m' h_{n,\beta}'\,dx>0,\\
K_2&\equiv\int_0^\infty \left(h_{n,\beta}'\right)^2dx>0.
\end{align}
The boundary terms vanish for even $m,n\ge2$. Therefore, for fixed $(m,n,\beta)$ there is exactly one positive deformation amplitude at which the global VIQ changes sign,
\begin{equation}
a_{\rm VIQ}
=\frac{J_1+\sqrt{J_1^2+4K_2J_0}}{2K_2},
\label{eq:aVIQ}
\end{equation}
with
\begin{equation}
a<a_{\rm VIQ}:\ I_{\rm VIQ}<0,
\qquad
a>a_{\rm VIQ}:\ I_{\rm VIQ}>0.
\end{equation}
At fixed dimensionless amplitude $a$, Eq.~\eqref{eq:IVIQ_quadratic} shows that $\alpha$ only multiplies $I_{\rm VIQ}$ by the positive factor $\alpha^2$, so it cannot change the sign or the location of the zero in $a$. At fixed dimensional amplitude $A$, however, $a=A/(\alpha r_0)$ varies with $\alpha$, and this conclusion no longer applies.

This should be contrasted with the quantity controlling the tangential NEC and the NED sign sector,
\begin{equation}
Q(x)=\alpha^2\left[(\mathcal R')^2+\mathcal R\mathcal R''\right]-1.
\label{eq:Q_dimensionless_IV}
\end{equation}
The boundary $Q_{\max}=0$ therefore depends explicitly on $\alpha$ and need not coincide with $I_{\rm VIQ}=0$. For the equal-power slice $(m,n)=(4,4)$, $\beta=1$, and $\alpha=0.5$, one finds
\begin{equation}
a_Q\simeq1.107434,
\qquad
a_{\rm VIQ}\simeq2.406874,
\end{equation}
corresponding respectively to $A_Q\simeq0.553717$ and $A_{\rm VIQ}\simeq1.203437$ when $r_0=1$. Thus the interval $a_Q<a<a_{\rm VIQ}$ already contains configurations in which $Q>0$ somewhere while the global VIQ remains negative. At $a=a_{\rm VIQ}$ the positive and negative local contributions compensate exactly, while local radial-NEC violation at the throats persists.

\begin{figure}[t]
    \centering
    \includegraphics[width=0.75\textwidth]{ 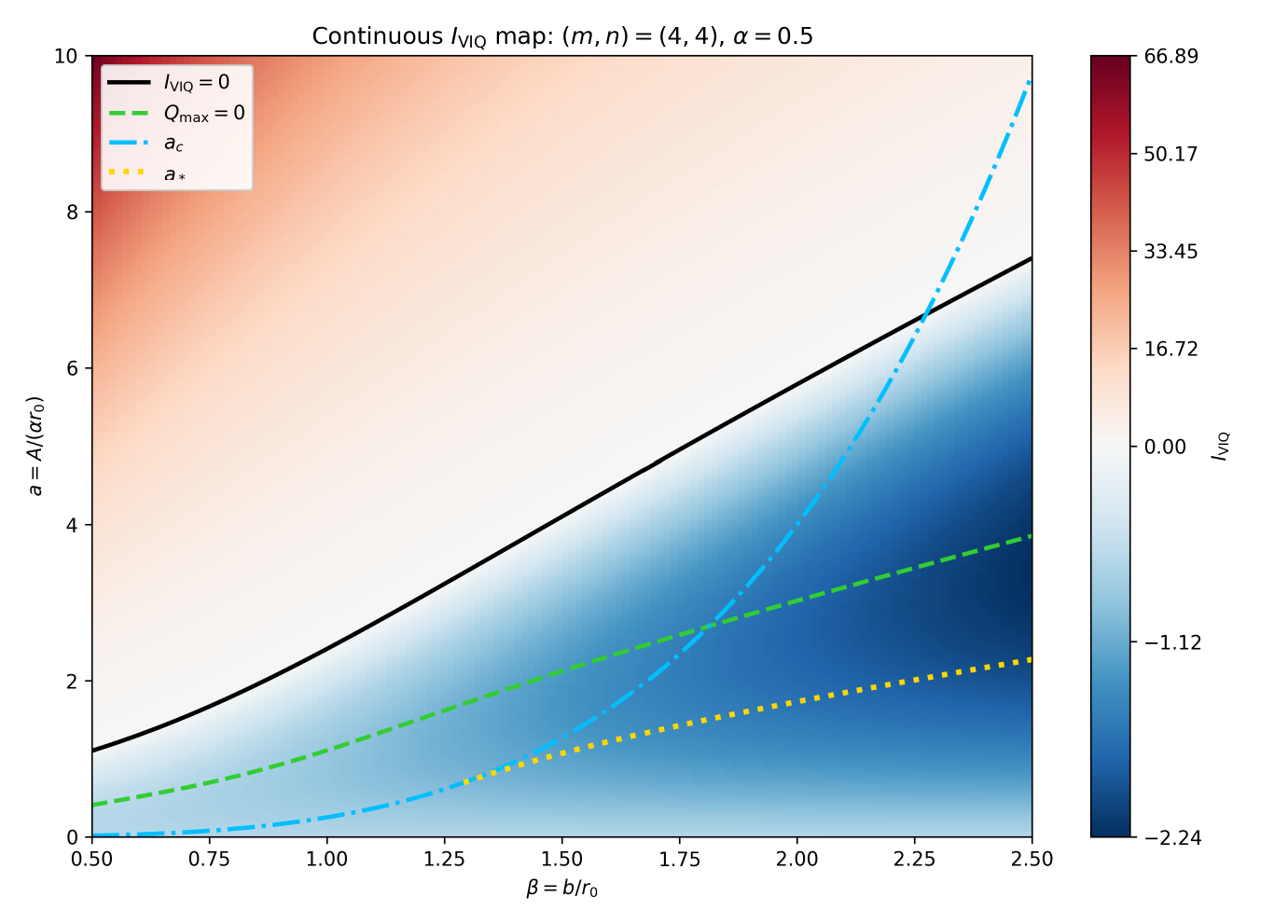}
    \caption{Continuous Nandi--Zhang--Kumar volume integral $I_{\rm VIQ}$ in the $(\beta,a)$ plane for the equal-power sector $(m,n)=(4,4)$ at $\alpha=0.5$. Blue shades correspond to $I_{\rm VIQ}<0$ and red shades to $I_{\rm VIQ}>0$; the black curve is $I_{\rm VIQ}=0$ and the green dashed curve is $Q_{\max}=0$. The cyan dash-dotted line is the central-transition curve $a_c=\beta^4/4$. For $\beta>\beta_c=(8/3)^{1/4}$, the yellow dotted curve $a_*$ supplies the additional global bifurcation boundary, with $a_*<a_c$. The color scale is evaluated with $r_0=1$ and $\kappa=8\pi$.}
    \label{fig:IVIQ_map_44}
\end{figure}

\begin{figure*}[t]
    \centering
    \includegraphics[width=0.75\textwidth]{ 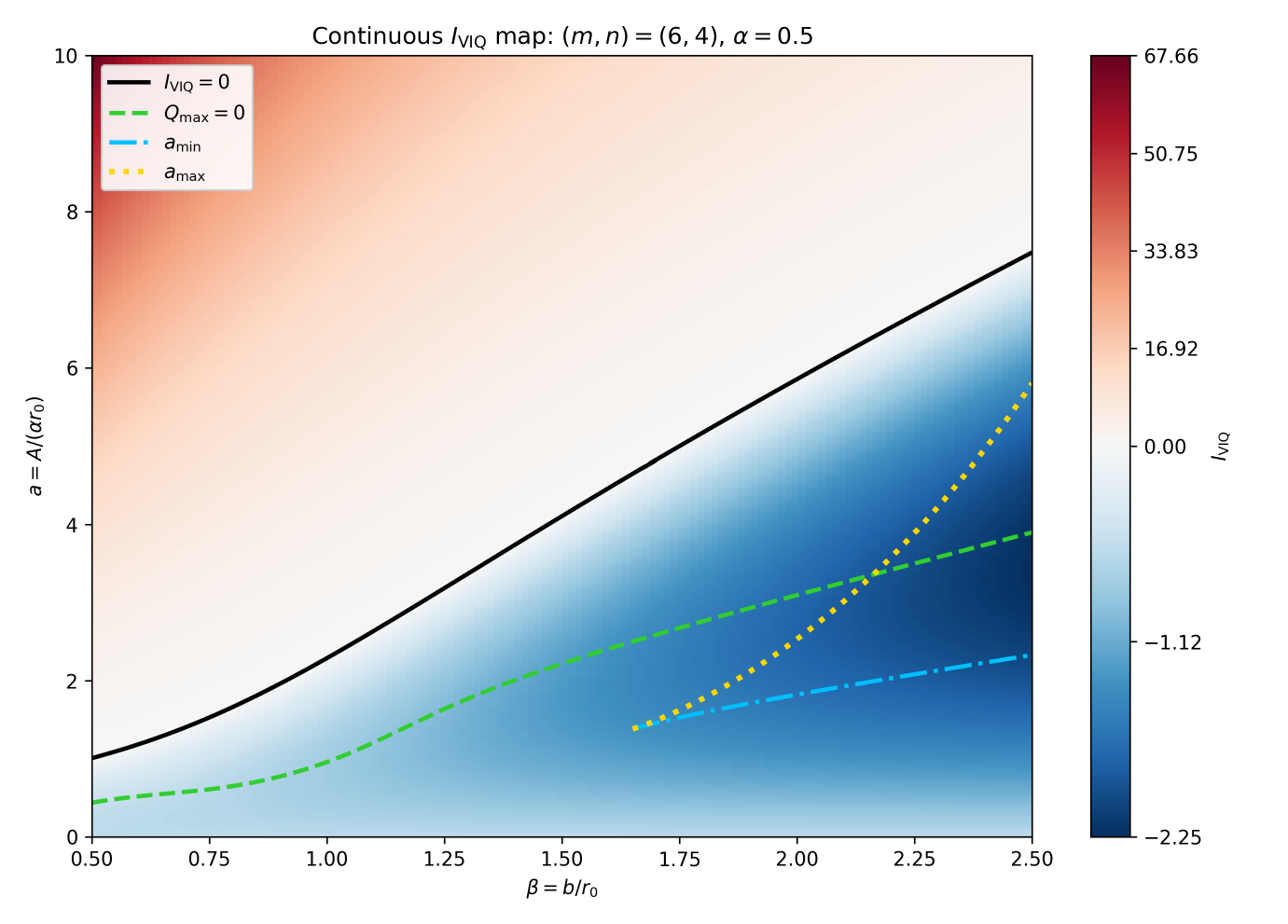}
    \caption{Continuous Nandi--Zhang--Kumar volume integral $I_{\rm VIQ}$ in the $(\beta,a)$ plane for $(m,n)=(6,4)$ at $\alpha=0.5$. The black and green dashed curves mark $I_{\rm VIQ}=0$ and $Q_{\max}=0$, respectively. The cyan dash-dotted and yellow dotted curves are the topological boundaries $a_{\min}$ and $a_{\max}$; they merge at $\beta_c\simeq1.650813$ and delimit the $4T$ window for larger $\beta$. The color scale is evaluated with $r_0=1$ and $\kappa=8\pi$.}
    \label{fig:IVIQ_map_64}
\end{figure*}

Figures~\ref{fig:IVIQ_map_44} and \ref{fig:IVIQ_map_64} show the continuous VIQ across the $(\beta,a)$ plane for two representative topological sectors, both at $\alpha=0.5$. The color scale gives the numerical value of $I_{\rm VIQ}$; the black curve marks $I_{\rm VIQ}=0$ and the green dashed curve marks $Q_{\max}=0$. The two boundaries are manifestly distinct. In both displayed parameter windows the $Q_{\max}=0$ curve is crossed first as $a$ increases, leaving a broad region with $Q>0$ somewhere but $I_{\rm VIQ}<0$ before the VIQ changes sign.

The $m>n$ sector provides a complementary example in which the topology itself develops a finite four-throat window. For $(m,n)=(6,4)$, the two global bifurcation curves $a_{\min}(\beta)$ and $a_{\max}(\beta)$ coalesce at
\begin{equation}
\beta_c\simeq1.650813,
\qquad
a_{\min}=a_{\max}\simeq1.382080,
\end{equation}
and split for $\beta>\beta_c$. The region $a_{\min}<a<a_{\max}$ is the $4T$ sector, while the neighboring deformed regions are $2T$. Figure~\ref{fig:IVIQ_map_64} shows that these topological boundaries cut across the continuous VIQ landscape rather than tracking either the VIQ-zero or $Q_{\max}=0$ curves. Hence throat multiplicity, the global sign of the integrated radial NEC, and the global sign of $Q$ are three logically distinct classifications of the same geometric family.

Although the two maps fix $\alpha=0.5$ for direct comparison, Eqs.~\eqref{eq:IVIQ_quadratic} and \eqref{eq:Q_dimensionless_IV} show that their relative placement changes with $\alpha$. In particular, positive global VIQ is compatible with globally negative $Q$. As an explicit example, $(m,n)=(4,4)$, $\beta=1$, $a=3$, and $\alpha=0.2$ gives, for $r_0=1$ and $\kappa=8\pi$,
\begin{equation}
I_{\rm VIQ}\simeq0.12387,
\qquad
Q_{\max}\simeq-0.33691.
\end{equation}
Thus the sign of the global VIQ cannot be inferred from the sign of $Q$, and conversely the appearance of a finite region with $Q>0$ does not determine the sign of the integrated radial NEC.

A hydrodynamic sound-speed diagnostic $v_s^2=p_r'/\rho'$ is not appropriate for this source. The canonical scalar, phantom scalar, and nonlinear electromagnetic field form a coupled field system, so $p_r'/\rho'$ is not the propagation speed of any of its perturbative degrees of freedom. Stability must instead be addressed through the coupled linearized field equations.

\section{Null and Timelike Geodesics and Light-Ring Structure}
\label{sec:geodesics}

The ultrastatic character of the geometry makes the geodesic structure especially transparent: for test particles minimally coupled to the background metric, the radial effective potentials are controlled entirely by the areal-radius profile $r(\ell)$. Wormhole light rings and their observational consequences have been studied extensively for Ellis-type and asymmetric geometries \cite{OhgamiSakai2015,Wielgus2020,HuangKunz2023}. Here the throat--belly structure derived in Sec.~\ref{field_eqs} can be translated directly into a hierarchy of radially unstable and stable circular orbits.

The NED source introduces one caveat. High-frequency electromagnetic disturbances in a nonlinear theory generally propagate on an effective optical geometry rather than on the null cone of the background metric \cite{Novello2000}. Since $\mathcal G_{FF}\neq0$ here, the orbits below describe minimally coupled massless test probes and will be referred to as \emph{background-metric} light rings. The physical electromagnetic characteristics of the source require the effective optical metric and, because the coupling depends on $\chi$, the coupled scalar--electromagnetic perturbation problem.

\subsection{Null geodesics and background-metric light rings}

Restricting, without loss of generality, to the equatorial plane $\theta=\pi/2$, the Killing symmetries yield the conserved energy and angular momentum of a massless test particle,
\begin{equation}
E=\dot t,
\qquad
L=r^2(\ell)\dot\varphi,
\end{equation}
where the dot denotes differentiation with respect to an affine parameter. The null constraint gives
\begin{equation}
\dot\ell^2+V_\gamma(\ell)=E^2,
\qquad
V_\gamma(\ell)=\frac{L^2}{r^2(\ell)}.
\label{eq:null_radial_potential}
\end{equation}
Circular null geodesics therefore satisfy
\begin{equation}
V_\gamma'(\ell_c)=0
\quad\Longleftrightarrow\quad
r'(\ell_c)=0.
\label{eq:light_ring_condition}
\end{equation}
Thus every strict extremum of the areal radius supports a background-metric circular null orbit. At a nondegenerate extremum,
\begin{equation}
V_\gamma''(\ell_c)
=-\frac{2L^2 r''(\ell_c)}{r^3(\ell_c)}.
\label{eq:light_ring_stability}
\end{equation}
Consequently, a throat ($r''>0$) corresponds to a maximum of $V_\gamma$ and hence to a radially unstable light ring, whereas a belly ($r''<0$) corresponds to a minimum and hence to a radially stable light ring.

At a higher-order stationary point Eq.~\eqref{eq:light_ring_stability} is inconclusive. Let the first nonvanishing correction be
\begin{equation}
r(\ell)=r_c+C_k(\ell-\ell_c)^k+\cdots,
\qquad k\geq3.
\end{equation}
Then
\begin{equation}
V_\gamma(\ell)=\frac{L^2}{r_c^2}
-\frac{2L^2 C_k}{r_c^3}(\ell-\ell_c)^k+\cdots.
\label{eq:degenerate_light_ring}
\end{equation}
For even $k$, the stationary point is a degenerate extremum: $C_k>0$ gives a throat and an unstable light ring, while $C_k<0$ gives a belly and a stable one. Odd $k$ gives a stationary inflection instead. At a generic noncentral saddle-node, $k=3$ and the coalesced circular orbit is marginal (one-sided, or semi-stable). These classifications refer to radial motion in the geodesic effective potential, not to dynamical stability of the spacetime. The central belly of the reference $m=n=4$ two-throat geometry provides an even-order example. Using
\begin{equation}
r(\ell)=\frac34-\frac18\ell^4+\mathcal O(\ell^8),
\end{equation}
one obtains
\begin{equation}
V_\gamma(\ell)=\frac{16L^2}{9}
+\frac{16L^2}{27}\ell^4+\mathcal O(\ell^8),
\label{eq:fiducial_quartic_light_ring}
\end{equation}
which is a radially stable degenerate minimum even though $V_\gamma''(0)=0$.

\begin{figure*}[t]
    \centering
    \includegraphics[width=\textwidth]{ 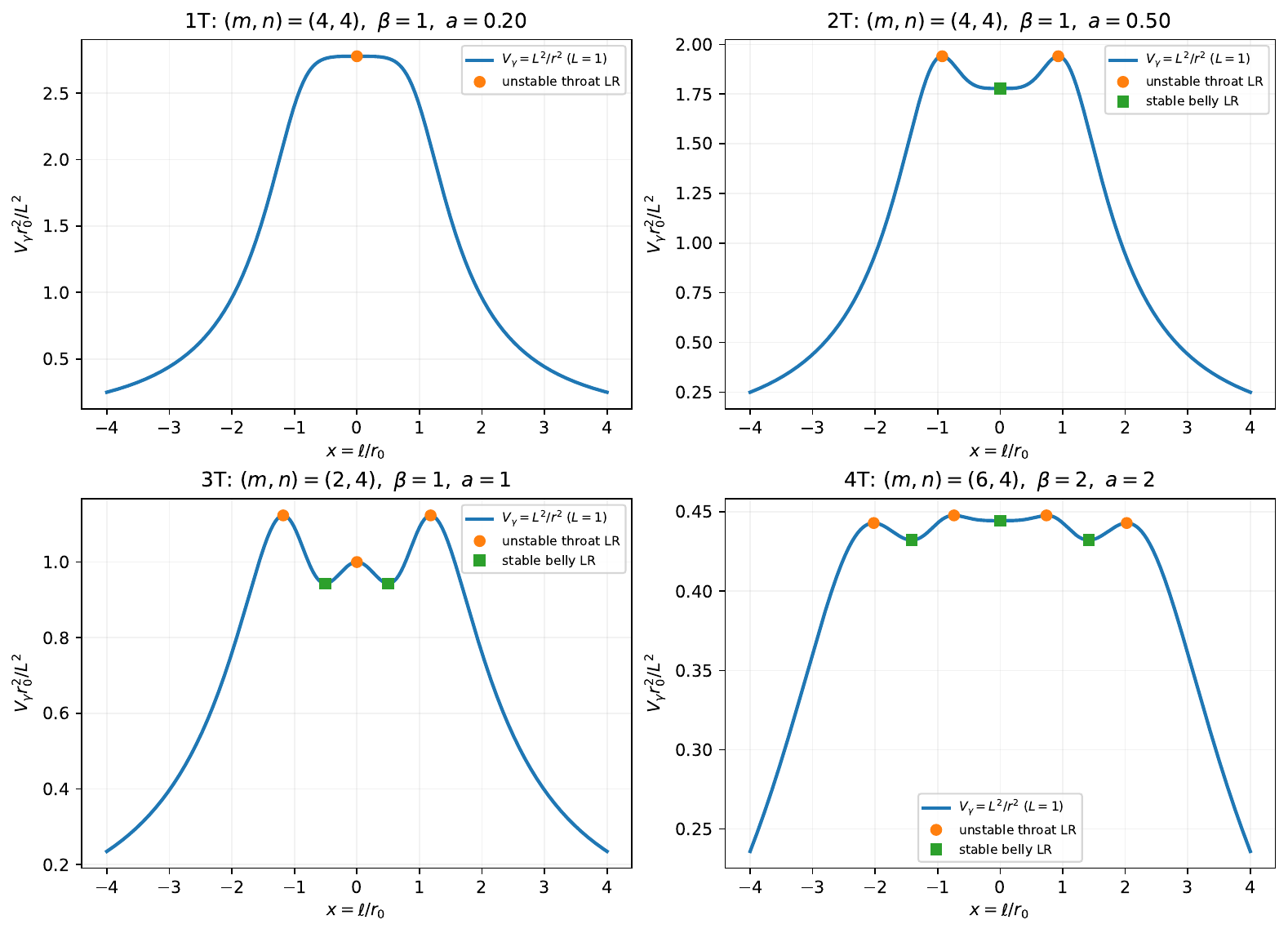}
    \caption{Background-metric null effective potential $V_\gamma=L^2/r^2$ for representative noncritical $1T$, $2T$, $3T$, and $4T$ geometries, with $L=1$. For these regular configurations, circular null orbits occur at extrema of $r(\ell)$: throat light rings are radially unstable maxima of $V_\gamma$ (circles), while belly light rings are radially stable minima (squares), including even-order degenerate extrema whose classification is determined by the first nonvanishing term. At a saddle-node boundary the coalesced orbit is instead a marginal stationary inflection, as discussed in the text. The geometric parameters are the same representative choices used in the topological survey.}
    \label{fig:null_potential_topologies}
\end{figure*}

Away from the bifurcation curves, the alternating sequence of throats and bellies implies a simple counting rule: a geometry with $N_T$ throats contains $N_T$ radially unstable throat light rings and $N_T-1$ radially stable belly light rings. Figure~\ref{fig:null_potential_topologies} illustrates this correspondence for representative $1T$, $2T$, $3T$, and $4T$ configurations. The single-throat geometry already possesses one unstable throat light ring; the new feature introduced by multi-throat topologies is the appearance of one or more stable internal light rings. At a noncentral fold, the coalescing stable/unstable pair is replaced exactly on the boundary by the marginal stationary-inflection orbit described above, so the regular counting rule applies only away from the bifurcation curves. Stable trapping and stable light rings in horizonless compact geometries are associated with very slow wave decay, long-lived modes, and possible nonlinear instabilities \cite{Cardoso2014StableLR,Keir2016,CunhaBertiHerdeiro2017}. Their appearance in the present family provides an additional reason to study the coupled perturbation spectrum.

\subsection{Conserved impact-parameter ratio and internal null trapping}

To avoid confusion with the geometric width parameter $b$, we define the conserved ratio
\begin{equation}
\mathfrak b_\gamma\equiv\frac{L}{E}.
\label{eq:conserved_impact_ratio}
\end{equation}
It is this quantity that enters the exact radial equation,
\begin{equation}
\frac{\dot\ell^2}{E^2}=1-\frac{\mathfrak b_\gamma^2}{r^2(\ell)}.
\label{eq:null_impact_equation}
\end{equation}
Because the spacetime is asymptotically conical, $\mathfrak b_\gamma$ should not be identified without qualification with the Euclidean impact parameter of an asymptotic observer. On the equatorial asymptotic cone,
\begin{equation}
ds^2\simeq-dt^2+d\ell^2+\alpha^2\ell^2d\varphi^2,
\end{equation}
and the locally unwrapped angular coordinate $\widetilde\varphi=\alpha\varphi$ has conjugate angular momentum $J=L/\alpha$. The corresponding asymptotic physical impact parameter is therefore
\begin{equation}
b_\infty\equiv\frac{J}{E}=\frac{\mathfrak b_\gamma}{\alpha}.
\label{eq:asymptotic_physical_impact}
\end{equation}
The internal turning-point analysis is most simply expressed in terms of the conserved ratio $\mathfrak b_\gamma$; the conversion in Eq.~\eqref{eq:asymptotic_physical_impact} may be applied when relating it to asymptotic measurements.

For the reference symmetric $2T$ geometry, let $r_T$ denote the throat radius and $r_B$ the central-belly radius. Numerically,
\begin{equation}
r_T\simeq0.717936,
\qquad
r_B=0.75.
\end{equation}
A background-metric null test particle sent from either asymptotic region then falls into three classes:
\begin{equation}
\begin{array}{lll}
\mathfrak b_\gamma<r_T &: & \text{transmission through both throats},\\
\mathfrak b_\gamma=r_T &: & \text{critical approach to the unstable throat light ring},\\
\mathfrak b_\gamma>r_T &: & \text{scattering back to the same asymptotic region}.
\end{array}
\label{eq:impact_classes}
\end{equation}
Equivalently, the asymptotic physical thresholds are obtained by dividing the right-hand-side radii by $\alpha$. There is therefore no capture channel for a conservative background-metric null geodesic: it is transmitted, scattered, or asymptotically approaches the critical orbit.

The central potential well nevertheless supports an internal trapped family. For
\begin{equation}
r_T<\mathfrak b_\gamma<r_B,
\label{eq:internal_trapping_band}
\end{equation}
a null geodesic initialized between the throats oscillates between two internal turning points around the stable central light ring. At $\mathfrak b_\gamma=r_B$ the motion reduces to the stable circular null orbit itself. Rays arriving from infinity cannot populate this band: the same condition $\mathfrak b_\gamma>r_T$ produces an outer turning point before the first throat is crossed. We call this family \emph{internal null trapping}. Its population depends on initial data inside the inter-throat region, so it does not define a capture channel for radiation incident from infinity.

\subsection{Timelike geodesics}

For a particle of rest mass $\mu$, define the specific energy and angular momentum
\begin{equation}
\varepsilon\equiv\frac{E}{\mu},
\qquad
j\equiv\frac{L}{\mu}.
\end{equation}
The equatorial radial equation becomes
\begin{equation}
\dot\ell^2+V_m(\ell)=\varepsilon^2,
\qquad
V_m(\ell)=1+\frac{j^2}{r^2(\ell)},
\label{eq:timelike_potential}
\end{equation}
where the dot now denotes differentiation with respect to proper time. For $j\neq0$, the circular-orbit condition is again
\begin{equation}
r'(\ell_c)=0,
\end{equation}
and
\begin{equation}
V_m''(\ell_c)
=-\frac{2j^2r''(\ell_c)}{r^3(\ell_c)}.
\end{equation}
Hence the locations and radial stability classification of timelike circular orbits coincide with those of the background-metric null light rings: throats support radially unstable circular orbits, while bellies support radially stable ones. Even-order central degeneracies and odd-order saddle-node folds inherit the same higher-order distinction described above; at a generic fold the circular orbit is marginal rather than a throat/belly extremum. Internal bounded timelike trajectories may likewise exist when initialized inside a potential well with an energy below the neighboring throat barriers.

The case $j=0$ is qualitatively different. Ultrastaticity makes $V_m=1$ constant, so worldlines at fixed spatial position are geodesic and no radial location is singled out as an equilibrium radius. The corresponding crossing time has the flat-form relation associated with the proper-radial coordinate and adds no independent characterization of the wormhole.

\section*{CONCLUSIONS}

We have constructed an ultrastatic generalized Ellis--Bronnikov family in which a localized deformation of the areal radius generates one-, two-, three-, and four-throat geometries. Solving the extremum problem globally gives the saddle-node curves for noncentral throat--belly creation and the separate central transition. The local character of the center is therefore insufficient to infer the global throat number. The entire family remains regular in the parameter domain studied: $r(\ell)$ stays positive, the curvature invariants are finite at ordinary and critical points, and the spacetime is geodesically complete. For $0<\alpha<1$ the two asymptotic ends are conical, with a finite solid-angle deficit.

The same geometry can be supported in ordinary General Relativity by a canonical scalar, a shift-symmetric phantom scalar, and a magnetic NED field. The scalar kinetic signs remain fixed; the relative gradients determine which field dominates the radial NEC combination. In the reference multi-throat solution, equal values of the magnetic invariant occur at points that require different $\mathcal G_F$, obstructing a global single-valued uncoupled NED. Coupling the electromagnetic sector to a monotonic scalar removes that ambiguity within the chosen reconstruction. For $\mathcal G=F+\lambda\sqrt F$, the NED contribution also carries the leading $r^{-2}$ asymptotic stress and approaches the radial string-cloud equation of state. This decomposition of the total stress is model dependent, as expected for an inverse reconstruction.

The pointwise energy conditions are tied closely to the throat--belly structure. A simple sufficient condition, $r''\leq0$ with $|r'|<1$, guarantees the WEC, DEC, and SEC for the total stress tensor; the representative 2T and 4T geometries contain finite regions of this type around their bellies. The phantom field remains active there and is compensated by the other sectors. Radial-NEC violation occurs at the throats and continues into integrable asymptotic tails. Its Nandi--Zhang--Kumar volume integral is exactly quadratic in the dimensionless amplitude $a$, with one positive zero for fixed $(m,n,\beta)$. The VIQ-zero curve, $Q_{\max}=0$, and the topological bifurcations are distinct boundaries, so the sign of the global integral does not determine the local throat violation.

For minimally coupled test particles, the circular-orbit structure follows the same geometry. Away from bifurcation curves, each throat carries a radially unstable light ring and each intermediate belly a stable one, giving $N_T$ unstable and $N_T-1$ stable light rings for a configuration with $N_T$ throats. A generic noncentral fold replaces the coalescing pair by a marginal stationary-inflection orbit. The reference 2T geometry also has a quartically degenerate stable central light ring and an internal trapped null family inaccessible to conservative rays from infinity. Timelike circular orbits with nonzero angular momentum obey the same radial classification. These statements concern the background metric; electromagnetic disturbances of the NED source generally require the corresponding effective optical geometry.

The remaining open problem is dynamical viability. The source is an inverse reconstruction, it is not unique, and it contains a genuine phantom field. Simpler ghost-scalar wormholes are known to be unstable \cite{GonzalezGuzmanSarbach2009}, while the stable trapping found here can support long-lived modes \cite{Cardoso2014StableLR,Keir2016,CunhaBertiHerdeiro2017}. The relevant test is therefore the coupled metric--scalar--electromagnetic perturbation problem. Once that sector is understood, the effective NED optical geometry can be used for ray tracing, lensing and critical-curve calculations under specified illumination conditions. Quasinormal spectra and the dynamical formation of the bifurcation pattern are natural extensions. The present results provide the exact background framework on which those questions can be posed.

\section*{Acknowledgments}
CRM would like to thank Conselho Nacional de Desenvolvimento Cient\'{i}fico e Tecnol\'ogico (CNPq) for the partial financial support, through grant 301122/2025-3. MSC and VHUB acknowledge support from Fundação Cearense de Apoio ao Desenvolvimento Científico e Tecnológico (FUNCAP) through contracts ICA-0252-00030.01.68/25 and BMD-0008-10054.01.06/26, respectively.

\bibliographystyle{apsrev4-2}
\bibliography{ref}

\end{document}